# NuCLS: A scalable crowdsourcing, deep learning approach and dataset for nucleus classification, localization and segmentation

Mohamed Amgad[1], Lamees A. Atteya[2]*, Hagar Hussein[3]*, Kareem Hosny Mohammed[4]*, Ehab Hafiz[5]*, Maha A.T. Elsebaie[6]*, Ahmed M. Alhusseiny[7], Mohamed Atef AlMoslemany[8], Abdelmagid M. Elmatboly[9], Philip A. Pappalardo[10], Rokia Adel Sakr[11], Pooya Mobadersany[1], Ahmad Rachid[12], Anas M. Saad[13], Ahmad M. Alkashash[14], Inas A. Ruhban[15], Anas Alrefai[12], Nada M. Elgazar[16], Ali Abdulkarim[17], Abo-Alela Farag[12], Amira Etman[8], Ahmed G. Elsaeed[16], Yahya Alagha[17], Yomna A. Amer[8], Ahmed M. Raslan[18], Menatalla K. Nadim[19], Mai A.T. Elsebaie[12], Ahmed Ayad[20], Liza E. Hanna[3], Ahmed Gadallah[12], Mohamed Elkady[21], Bradley Drumheller[22], David Jaye[22], David Manthey[23], David A. Gutman[24], Habiba Elfandy[25, 26], Lee A.D. Cooper[1, 27, 28]

\* These authors contributed equally

[1] Department of Pathology, Northwestern University, Chicago, IL, USA, [2] Cairo Health Care Administration, Egyptian Ministry of Health, Cairo, Egypt, [3] Department of Pathology, Nasser institute for research and treatment, Cairo, Egypt, [4] Department of Pathology and Laboratory Medicine, University of Pennsylvania, PA, USA, [5] Department of Clinical Laboratory Research, Theodor Bilharz Research Institute, Giza, Egypt, [6] Department of Medicine, Cook County Hospital, Chicago, IL, USA, [7] Department of Pathology, Baystate Medical Center, University of Massachusetts, Springfield, MA, USA, [8] Faculty of Medicine, Menoufia University, Menoufia, Egypt, [9] Faculty of Medicine, Al-Azhar University, Cairo, Egypt, [10] Consultant for The Center for Applied Proteomics and Molecular Medicine (CAPMM), George Mason University, Manassas, VA, USA, [11] Department of Pathology, National Liver Institute, Menoufia University, Menoufia, Egypt, [12] Faculty of Medicine, Ain Shams University, Cairo, Egypt, [13] Cleveland Clinic Foundation, Cleveland, OH, USA, [14] Department of Pathology, Indiana University, Indianapolis, IN, USA, [15] Faculty of Medicine, Damascus University, Damascus, Syria, [16] Faculty of Medicine, Mansoura University, Mansoura, Egypt, [17] Faculty of Medicine, Cairo University, Cairo, Egypt, [18] Department of Anaesthesia and Critical Care, Menoufia University Hospital, Menoufia, Egypt, [19] Department of Clinical Pathology, Ain Shams University, Cairo, Egypt, [20] Research Department, Oncology Consultants, PA, Houston, TX, USA, [21] Siparadigm Diagnostic Informatics, Pine Brook, NJ, USA, [22] Department of Pathology and Laboratory Medicine, Emory University School of Medicine, Atlanta, GA, USA, [23] Kitware Inc., Clifton Park, NY, USA, [24] Department of Neurology, Emory University School of Medicine, Atlanta, GA, USA, [25] Department of Pathology, National Cancer Institute, Cairo, Egypt, [26] Department of Pathology, Children's Cancer Hospital Egypt (CCHE 57357), Cairo, Egypt, [27] Lurie Cancer Center, Northwestern University, Chicago, IL, USA, [28] Center for Computational Imaging and Signal Analytics, Northwestern University Feinberg School of Medicine, Chicago, IL, USA

# Abstract

High-resolution mapping of cells and tissue structures provides a foundation for developing interpretable machine-learning models for computational pathology. Deep learning algorithms can provide accurate mappings given large numbers of labeled instances for training and validation. Generating adequate volume of quality labels has emerged as a critical barrier in computational pathology given the time and effort required from pathologists. In this paper we describe an approach for engaging crowds of medical students and pathologists that was used to produce a dataset of over 220,000 annotations of cell nuclei in breast cancers. We show how suggested annotations generated by a weak algorithm can improve the accuracy of annotations generated by non-experts and can yield useful data for training segmentation algorithms without laborious manual tracing. We systematically examine interrater agreement and describe modifications to the MaskRCNN model to improve cell mapping. We also describe a technique we call Decision Tree Approximation of Learned Embeddings (DTALE) that leverages nucleus segmentations and morphologic features to improve the transparency of nucleus classification models. The annotation data produced in this study are freely available for algorithm development and benchmarking at: https://sites.google.com/view/nucls.

---

# Introduction

## Motivation

Convolutional neural networks (CNN) and other deep learning methods have been at the heart of recent advances in medicine (see Table S1 for terminology) [1]. A key challenge in computational pathology is the scarcity of large-scale labeled datasets for model training and validation [2–4]. Specifically, there is a shortage of annotation data for delineating tissue regions and cellular structures in histopathology. This information is critical for training interpretable deep-learning models, as they allow the detection of concepts that map to known diagnostic criteria [4–7]. Moreover, the availability of tissue and region annotations enables high-resolution spatial mapping of the tumor microenvironment, paving the way for computationally-driven discovery of histopathologic biomarkers and biological associations [4,8–12]. This shortage is often attributed to the domain expertise required to produce annotation labels; with pathologists spending years in residency and fellowship training [2,13]. This is exacerbated by the time constraints of clinical practice, and the repetitive nature of annotation work. Manual tracing of object boundaries is an especially demanding task, and there is a pressing need to obtain this data using facilitated or assisted annotation strategies [14]. By comparison, traditional annotation problems like detection of people in natural images require almost no training and typically engage the general public [14]. Moreover, unique problems often require new annotation data, underscoring the need for scalable and reproducible annotation workflows [15].

We address these issues using an assisted annotation method that leverages the participation of non-pathologists (NPs), including medical students and graduates. Medical students typically have strong incentives to participate in annotation studies, with increased reliance on research participation in residency selection [16]. We describe adaptations to both the data collection and computational modeling aspects to improve scalability and to reduce effort. This work focuses on nucleus classification, localization and segmentation (NuCLS, for short) in whole-slide scans of hematoxylin and eosin-stained (H&E) slides of breast carcinoma from 18 institutions from The Cancer Genome Atlas (TCGA). Our annotation pipeline enables low-effort collection of nucleus segmentation and classification data, paving the way for systematic discovery of histopathologic-genomic associations and morphological biomarkers of disease progression [4,5,9,11,12].

**Previous work**

There has been growing interest in addressing data scarcity in histopathology by either scaling data generation or reducing reliance on labeled data [17–21]. This work is meant to fit into the broad context of scalable data generation in biomedical imaging domains where expert annotation is expensive and/or difficult. *Crowdsourcing*, the process of engaging a "crowd" of individuals to annotate data, is critical to solving this problem. There exists a large body of relevant work in crowdsourcing for medical image analysis [14,22,23]. Previously, we published a study and dataset using crowdsourcing of NPs for annotation of low-power regions in breast cancer [24]. Our approach was *structured*, in the sense that we assigned different tasks depending on the level of expertise, and leveraged collaborative annotation to obtain data that is large in scale, but also high in quality. Here, we significantly expand this idea by focusing on the challenging problems of nucleus classification, localization, and segmentation. This computer vision problem is a subject of significant interest in computational pathology [25–27].

While the public release of data is only one aspect of our study, it is important to note that there are related public datasets that can be used in conjunction with ours [26,28–31]. None, however, have systematically explored the data generation process. Annotation of stromal tumor-infiltrating lymphocytes (sTILs) is the subject of an ongoing study by the US Food and Drug Administration [32]. More generally, most public computational pathology datasets are either limited in scale, were generated through exhaustive annotation efforts by practicing pathologists, or do not disclose or discuss data generation [2,22,33]. Additionally, to the best of our knowledge, most other works do not explore interrater agreement (especially for experts vs non-experts) or do not provide solutions to adapt object detection frameworks for nucleus detection.

Two works are of particular relevance to this paper. A study by Irshad *et al* showed that non-experts, recruited through the Figure Eight platform, can produce accurate nucleus detections and segmentations in renal clear cell cancer, but was limited to 10 whole-slide images [19]. Recent work by Hou *et al* explored the use of synthetic data to produce nuclear segmentations. Their work, while an important contribution, did not address classification, relied on qualitative slide-level evaluations of results, and did not explore how algorithmic bias affects data quality [34].

**Our contributions**

In this work, we describe a scalable crowdsourcing approach that engaged NPs in a systematic way and produced annotations for localization, segmentation, and classification of nuclei in breast cancer. We obtained a total of 222,396 annotations. These include over 125,000 single-rater nucleus annotations and over 97,000 multi-rater annotations. This workflow required minimal effort from pathologists, and used algorithmic suggestions to scale the annotation process and to obtain *hybrid* annotation datasets containing a large number of segmentation boundaries without laborious manual tracing. We show that algorithmic suggestions can improve the accuracy of NP annotations and that NPs are reliable annotators of common cell types. We discuss a new constrained clustering method that we developed for reliable truth inference in multi-rater datasets. We also show how multi-rater data can be used to ensure the quality of NP annotations, or to replace expert supervision in some contexts.

Additionally, we show that MaskRCNN, the state-of-the-art object detection model, can be fundamentally modified for the specific task of nucleus detection and to learn from hybrid annotation datasets. We also describe a technique we call Decision Tree Approximation of Learned Embeddings (DTALE) that improves model explainability, addressing a barrier to clinical adoption of deep-learning methods. Finally, the annotation datasets we produced have been deposited at https://sites.google.com/view/nucls and can be used for model development and benchmarking.

## Results & discussion

### Structured crowdsourcing enables scalable data collection

Pathologist time is limited and expensive, and relying solely on pathologists for generating annotations can hinder development of state-of-the-art models based on CNNs. In this study, we show that NPs can perform most of the time-consuming annotation tasks, and that pathologist involvement can be limited to low-effort tasks that include:

1. Training NPs and answering their questions (Figure 1).
2. Qualitative scoring of NP annotations (Figure S1).
3. Low-power annotation of histologic regions (Figure S2) [24].

We used a web-based annotation platform called HistomicsUI for annotation, feedback, and quality review [35]. HistomicsUI provides a user interface with annotation tools and an API for programmatic querying and manipulation of the centralized annotation database. We obtained annotations from 32 NPs and 7 pathologists, located in the US, Egypt, Syria, Australia, and Maldives. We obtained 128,000 nucleus annotations from 3,944 fields-of-view (FOV) and 125 triple-negative breast cancer patients. The annotations included bounding box placement, classification, and for a sizable fraction of nuclei, segmentation boundaries. Half of these annotations underwent quality control correction based on feedback by a practicing pathologist.

Additionally, we obtained three multi-rater datasets containing 97,300 annotations, where the same FOV was annotated by multiple participants (Figure 1b, Figure 2). Collection of multi-rater data enables quantitative evaluation of NP reliability, interrater variability, and the impact of algorithmic suggestions on NP accuracy. Multi-rater annotations were not corrected by pathologists and enable unbiased assessment of NP performance. Pathologist annotations were also collected for a limited set of multi-rater FOVs to evaluate NP accuracy.

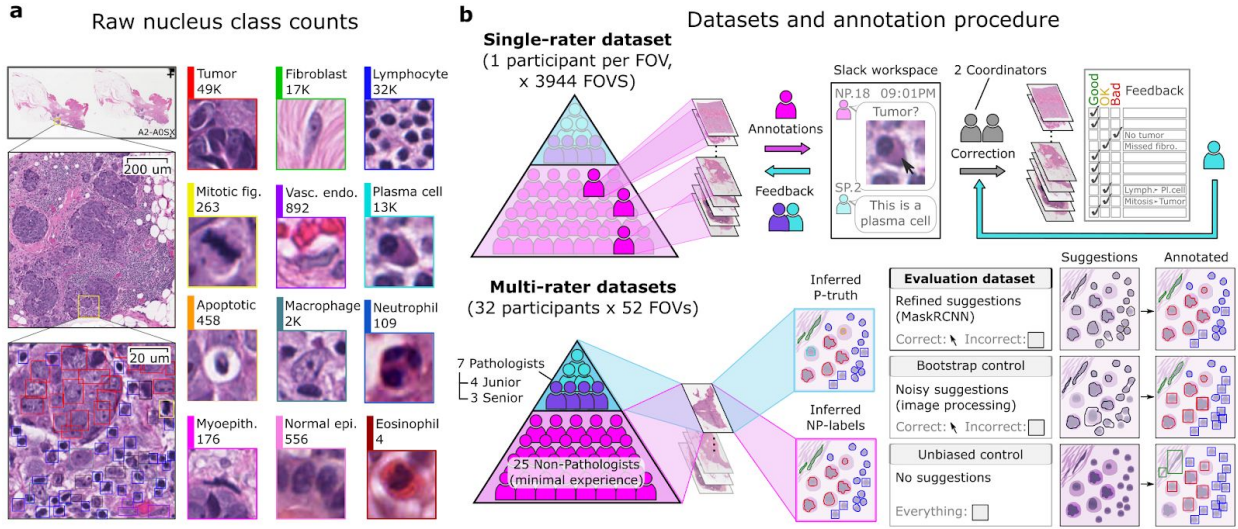

**Figure 1. Dataset annotation and quality control procedure. a.** Nucleus classes annotated. **b.** Annotation procedure and resultant datasets. Two approaches were used to obtain nucleus labels from non-pathologists (NPs). (Top) The first approach focused on breadth, collecting single-rater annotations over a large number of FOVs to obtain the majority of data in this study. NPs were given feedback on their annotations, and two study coordinators corrected and standardized all single-rater NP annotations based on input from a senior pathologist. (Bottom) The second approach focused on evaluating interrater reliability and agreement, obtaining annotations from multiple NPs for a smaller set of shared FOVs. Annotations were also obtained from pathologists for these FOVs to measure NP reliability. The procedure for inferring a single set of labels from multiple participants is described in figure 2, and we make a distinction between inferred NP-labels and inferred P-truth for clarity. Three multi-rater datasets were obtained: an Evaluation dataset, which is the main multi-rater dataset, as well as Bootstrap and Unbiased experimental controls to measure the value of algorithmic suggestions. In all datasets except the Unbiased control, participants were shown algorithmic suggestions for nucleus boundaries and classes. They were directed to click nuclei with correct boundary suggestions and annotate other nuclei with bounding boxes. The pipeline to obtain algorithmic suggestions consisted of two steps: 1. Using image processing to obtain bootstrapped suggestions. This corresponds to the Bootstrap control; 2. Training a MaskRCNN model to refine the bootstrapped suggestions. This corresponds to the single-rater and Evaluation datasets.

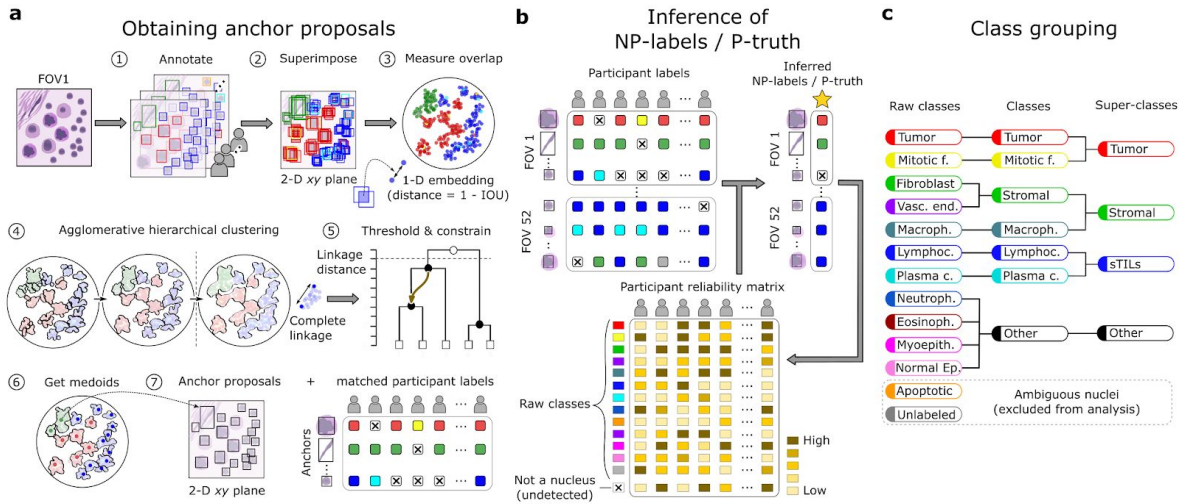

**Figure 2. Inference from multi-rater datasets.** The purpose of this step was to infer the nucleus locations and classifications from multi-rater data. **a.** The first step involved agglomerative hierarchical clustering of bounding boxes using Intersection-Over-Union (IOU) as a similarity measure. We imposed a constraint during clustering that prevents merging of annotations where a single participant has annotated overlapping nuclei. Participant intention was preserved by demoting annotations from the same participant to the next node (step 5, arrow). After clustering was complete, a threshold IOU value was used to obtain the final clusters (step 5, black nodes). Within each cluster, the medoid bounding box was chosen as an *anchor proposal*. The end result was a set of anchors with corresponding clustered annotations. When a participant did not match to an anchor, it was considered a conscious decision to not annotate a nucleus at that location. **b.** Once anchors were obtained, an expectation-maximization (EM) procedure was used to estimate: 1. which anchors represent actual nuclei, and 2. which classes to assign these anchors. The EM procedure estimates and accounts for the reliability of each participant for each classification. EM was performed separately for NPs and pathologists. **c.** Grouping of nucleus classes. Consistent with standard practice in object detection, nuclei were grouped, based on clinical reasoning, into five classes and three super-classes.

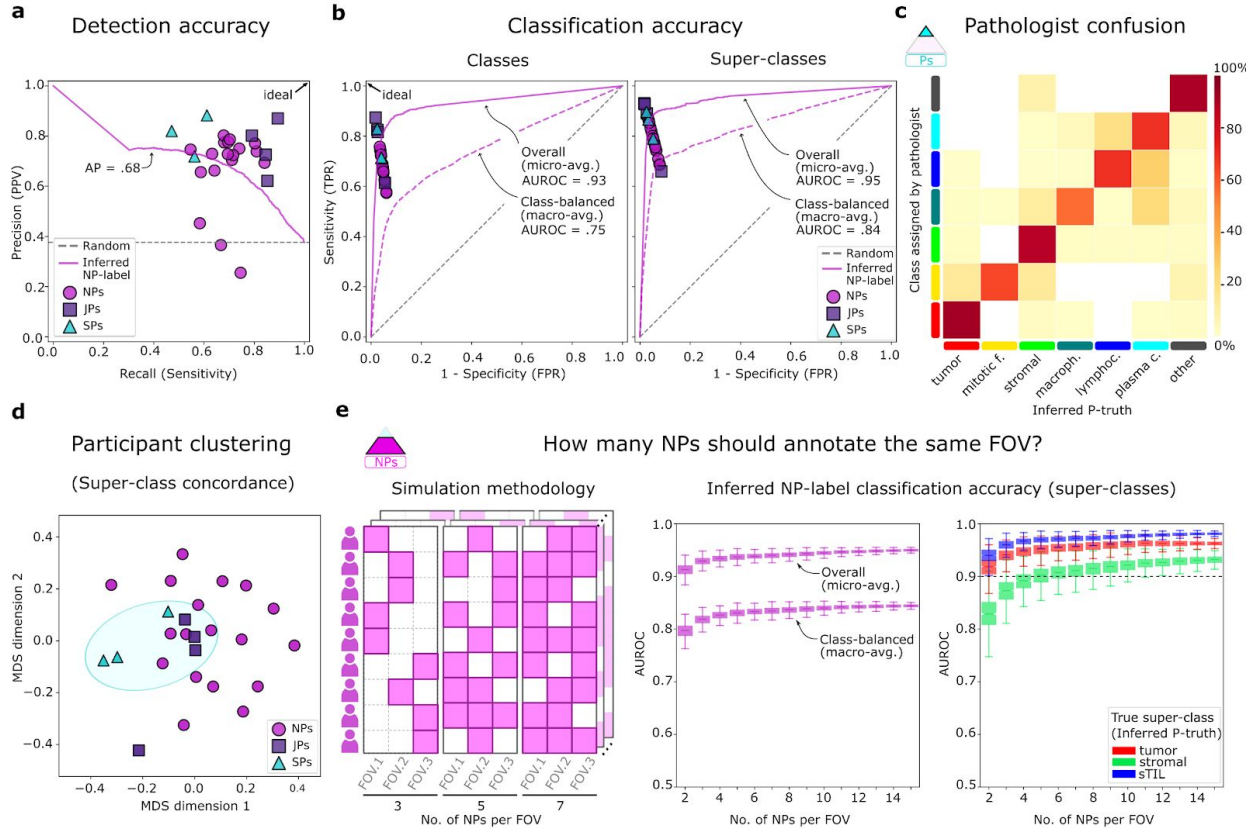

**Figure 3. Accuracy of participant annotations. a.** Detection precision-recall comparing annotations to inferred P-truth. Junior pathologists tend to have similar precision but higher recall than senior pathologists, possibly reflecting the time constraints of pathologists. **b.** Classification ROC for classes and super-classes. The overall classification accuracy of inferred NP-labels was high. Class-balanced accuracy (macro-average) is notably lower since NPs are less reliable annotators of uncommon classes. **c.** Confusion between pathologist annotations and inferred P-truth. **d.** Multidimensional scaling (MDS) analysis of interrater classification agreement. There is some clustering by participant experience (blue ellipse), highlighting the importance of modeling reliability during label inference. **e.** A simulation was used to measure how redundancy impacts the classification accuracy of inferred NP-labels. While keeping the total number of NPs constant, we randomly kept annotations for a variable number of NPs per FOV. Accuracy in these simulations was class-dependent, with stromal nuclei requiring more redundancy for accurate inference.

## NPs can reliably classify common cell types

Detection accuracy of NPs was moderately high (AP=0.68) and was similar to the detection accuracy of pathologists. Classification accuracy of NPs, on the other hand, was only high for common nucleus classes (AUROC=0.93 micro-average vs. 0.75 macro-average) and was higher when grouping by super-class ([Figure 3](#), [Figure S3](#)). We reported the same phenomenon in our previous work on crowdsourcing annotation of tissue regions [24]. We observed moderate clustering by participant experience ([Figure 3d](#)), and variability in classification accuracy among NPs (MCC=60.7-84.2). This motivated our quality control procedures. For the single-rater dataset, study coordinators manually corrected missing or misclassified cells, and practicing pathologists supervised and approved annotations. For the multi-rater datasets, we inferred a singular label from pathologists (P-truth) and NPs (NP-label) using an Expectation-Maximization (EM) framework that estimates reliability values for each participant [36,37].

When pathologist supervision is not an option, multi-rater datasets need to have annotations from a sufficient number of NPs to infer reliable data. We used the annotations we obtained to perform simulations to estimate the accuracy of inferred NP-labels with fewer numbers of participating NPs (Figure 3e). The inferred NP-label accuracy increased up to six NPs per FOV, after which there were diminishing returns. Our simulations also showed that stromal nuclei require more NPs per FOV than tumor nuclei or sTILs.

**Minimal-effort collection of nucleus segmentation data**

Many nucleus detection and segmentation algorithms were developed using conventional image analysis methods prior to widespread adoption of CNNs. These algorithms have little or no dependence on annotations, and while they may not be as accurate as CNNs, they can correctly segment a significant fraction of nuclei. We used simple nucleus segmentation heuristics combined with low-power region annotations to obtain *bootstrapped* annotation suggestions for nuclei (Figure S2). The suggestions were refined using a deep-learning model (MaskRCNN) as a function approximator trained on the bootstrapped suggestions. This allowed poor quality bootstrapped suggestions in one FOV to be smoothed by better suggestions in other FOVs (Figure S4, Table S2) and is analogous to fitting a regression line to noisy data [17,38]. This model was applied to the FOVs to generate refined suggestions that were shown to participants when annotating the single-rater dataset and the *Evaluation dataset* (the primary multi-rater dataset) [39]. Two additional multi-rater datasets were obtained as controls:

1. *Bootstrap control:* participants were shown unrefined bootstrapped suggestions.
2. *Unbiased control*: participants were not shown any suggestions. This was the first multi-rater dataset to be annotated.

Accurate suggestions can be confirmed during annotation with a single click, reducing effort and providing valuable nucleus boundaries that can aid development of segmentation models. Other nuclei can be annotated by participants with bounding boxes, that require more effort than click annotations but less effort than manual tracing [14]. We obtained a substantial proportion of nucleus boundaries through clicks: 41.7±17.3% for the Evaluation dataset and 36.6% for the single-rater dataset (Figure 4, Figure S5). The resultant hybrid dataset contained a mixture of bounding boxes and accurate segmentation boundaries (Evaluation dataset DICE=85.0±5.9). We argue that it is easier to handle hybrid datasets at the level of algorithm development than to have participants trace missing boundaries or correct imprecise ones. We evaluate the bias of using these suggestions in the following section.

**Algorithmic suggestions improve classification accuracy**

There was value in providing the participants with suggestions for nuclear class, which included suggestions directly inherited from low-power region annotations, as well as high-power refined suggestions produced by MaskRCNN (Figure 4). Pathologists had substantial self-agreement when annotating FOVs with or without refined suggestions (Kappa=87.4±7.9). NPs also had high self-agreement, but were more impressionable when presented with suggestions (Kappa=74.0±12.6). This was, however, associated with a reduction in bias in their annotations; refined suggestions improved the classification accuracy of inferred NP-labels (AUROC=0.95 vs. 0.92, p<0.001). This is consistent with Marzahl *et al*, who reported similar findings in a crowdsourcing study using bovine cytology slides [23].

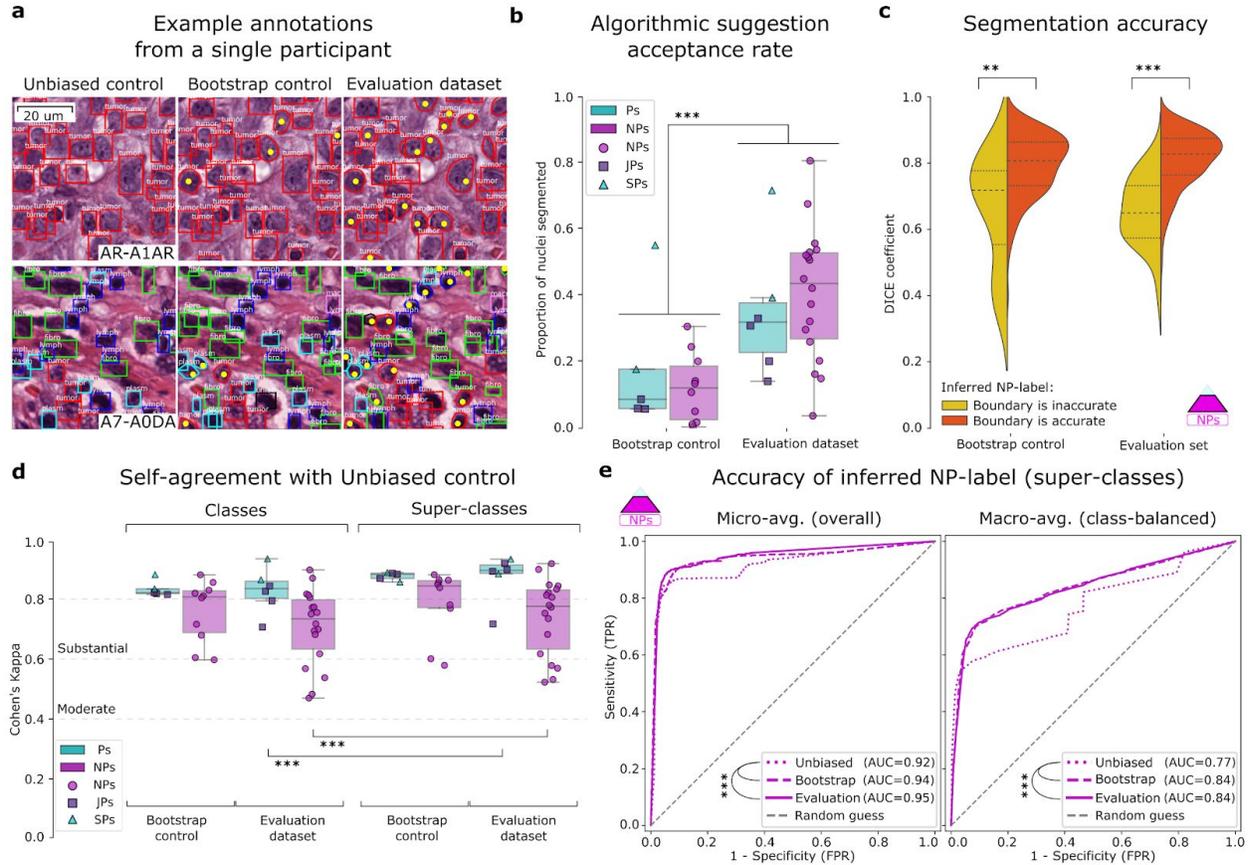

**Figure 4. Effect of algorithmic suggestions on annotation abundance and accuracy.** We compared annotations from the Evaluation dataset and controls to measure the impact of suggestions and MaskRCNN refinement on the acquisition of nucleus segmentation data and the accuracy of annotations. **a.** Example annotations from a single participant. Algorithmic suggestions allow collection of accurate nucleus segmentations without added effort. Yellow points indicate clicks to approve suggestions. **b.** The number of segmented nuclei clicked is significantly higher for the Evaluation dataset than for the Bootstrap control, indicating that refinement improves suggestion quality. **c.** Accuracy of algorithmic segmentation suggestions. The comparison is made against a limited set of manually traced segmentation boundaries obtained from one senior pathologist. Suggestions that were determined to be correct by the EM procedure had significantly more accurate segmentation boundaries. **d.** Self-agreement for annotations in the presence or absence of algorithmic suggestions. Agreement is substantial for both NP and pathologist groups, indicating that algorithmic suggestions do not adversely impact classification decisions. Pathologists have higher self-agreement and are less impressionable than NPs. **e.** ROC curves for the classification accuracy of inferred NP-label, using inferred P-truth as our reference. Statistically significant comparisons are indicated with a star (**, p<0.01; ***, p<0.001).

Region-based class suggestions for nuclei were, overall, more concordant with the corrected single-rater annotations compared to MaskRCNN refined (high-power) nucleus suggestions (MCC=67.6 vs. 52.7) ([Figure S4](#), [Table S2](#)). Nonetheless, high-power nucleus suggestions were more accurate for 24.8% of FOVs and had higher recall for sTILs (96.8 vs. 76.6) [4,12]. This makes sense, since stromal regions often contain scattered sTILs, and a region-based approach to labeling would incorrectly mark these as stromal nuclei (e.g. see [Figure S6](#)) [24,40]. Hence, the value of low and/or high-power classification suggestions is context dependent.

**Exploring nucleus detection and classification tradeoffs**

Naturally, there is some variability in the judgements made by participants about nuclear locations and classes, as well as the accuracy of suggested boundaries. We study the process of inferring a single truth from multi-rater datasets and discuss the effect of various parameters.

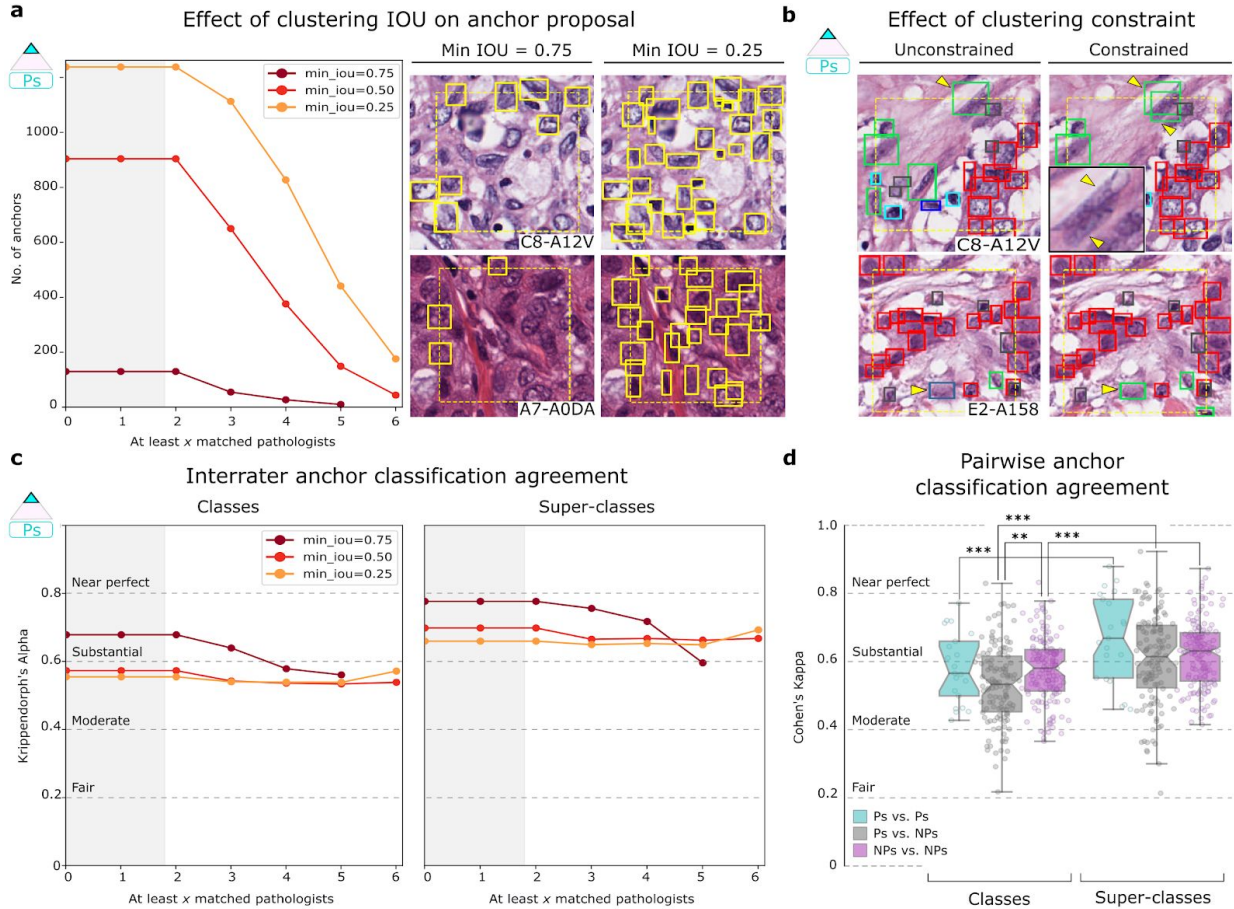

**Figure 5. Effect of clustering on detection and interrater agreement. a.** Stricter IOU thresholds reduce the number of anchor proposals generated by clustering but increase agreement. A threshold of 0.25 provides more anchor proposals with negligible difference in agreement from the 0.5 threshold. The shaded region indicates that by design there are no anchor proposals with less than two clustered annotations. **b.** The clustering constraint prevents annotations from the same participant from being assigned to the same anchor, preserving participant intention when annotating overlapping nuclei. This results in better detection of overlapping nuclei during clustering (upper panel) and also impacts the inferred P-truth for anchors (bottom panel). **c.** Interrater classification agreement among pathologists for tested clustering thresholds. **d.** Pairwise interrater classification agreement (Cohen's Kappa) at 0.25 IOU threshold. Statistically significant comparisons are indicated with a star (\*\*, p<0.01; \*\*\*, p<0.001).

There is a tradeoff between the number of nucleus anchor proposals and interrater agreement ([Figure 5](#)). The clustering IOU threshold that defines the minimum acceptable overlap between any two annotations had a strong impact on the number of anchor proposals. We found an IOU threshold of 0.25 detects most nuclei with adequate pathologist classification agreement (1,238 nuclei, Alpha=55.5). During clustering, we imposed a constraint to prevent annotations from the same participant from mapping to the same cluster. This improved detection of touching nuclei when the number of participants was limited as it was for pathologists ([Figure 5b](#)).

Nucleus detection was a larger source of discordance among participants than nucleus classification ([Figure 3](#), [Figure S7](#)). Some nucleus classes were easier to detect than others. sTILs were the easiest to detect, likely due to their hyperchromicity and tendency to aggregate; 53.3% of sTILs were detected by 16+ NPs ([Figure S8](#)). Fibroblasts were demonstrably harder to detect (only 21.4% were detected by 16+ NPs), likely because of their relative sparsity and lighter nuclear staining. Lymphocytes and plasma cells, which often co-aggregate in lymphoplasmacytic clusters, were a source of interrater discordance for both pathologists and NPs [4,41]. This may stem from variable degrees of reliance on low-power vs high-power morphologic features. Interrater agreement for nuclear classification was high, and significantly improved when classes were grouped into clinically-salient super-classes (Alpha=66.1 (pathologists) and 60.3 (NPs); Figure 5).

## NuCLS: a MaskRCNN variant using hybrid annotation datasets

Nucleus detection differs from natural object detection tasks in several important respects. Nuclei have lower variability in size and coarse morphology than objects in natural scenes, and different nucleus classes are mostly distinguished by fine detail and spatial context. Models designed for detection in natural images, including MaskRCNN, produce inferences that integrate the concepts of detection and classification (e.g. Person, 82% probability) [39]. In contrast, for the purpose of detection, nuclei belong to a single meta-class with an ovoid morphology. Treating nuclei as a single meta-class allows calculation of a full classification probability vector for each nucleus, which would be useful where nuclear morphology is ambiguous, especially in computer-assisted diagnostic settings. Nuclei are also typically much smaller and more numerous than natural objects, even at high magnification, which makes accurate detection more challenging (Figure S9) [42]. Moreover, scalable deployment of trained nucleus detection models requires the flexibility to perform inference for very large images without resizing and distorting nuclei [43,44].

We modified MaskRCNN for the specific task of nucleus detection and to handle the hybrid annotations generated by our assisted annotation method (Figure 6). Our key modifications included increased independence of the jointly-trained detector and classifier, and enabled: 1. training with hybrid box/segmentation annotations; 2. generating class probability vectors for all detections; 3. inference with variable input image sizes without distortion of scale or aspect ratios. To account for the scale and density of nuclei we also made the following changes to improve detection performance: 1. increasing the density of region proposals relative to natural image datasets; 2. digitally increasing magnification beyond 40x objective (Table S3). Since detection and classification have disparate clinical utility, we report their accuracies separately.

## Trained NuCLS models have high generalization accuracy

NuCLS models were trained on the corrected single-rater dataset, and reached convergence within 40 epochs (Figure S10). They converged smoothly despite being trained using a mixture of box and segmentation annotations. We used an internal-external cross-validation scheme to assess the generalization performance of our trained models (Figure 6c). This separates training and testing data by hospital rather than image to better reflect the challenge of external generalization [4,45]. Trained NuCLS models had high generalization accuracy for detection (AP=74.8±0.5), segmentation (DICE=88.5±0.8), and super-class classification (AUROC=93.5±2.7) (Table 1). For sTILs classification, NuCLS models had a testing AUROC of 94.7±2.1 (Table S4). This was also reflected on qualitative examination of predictions (Figure 6d).

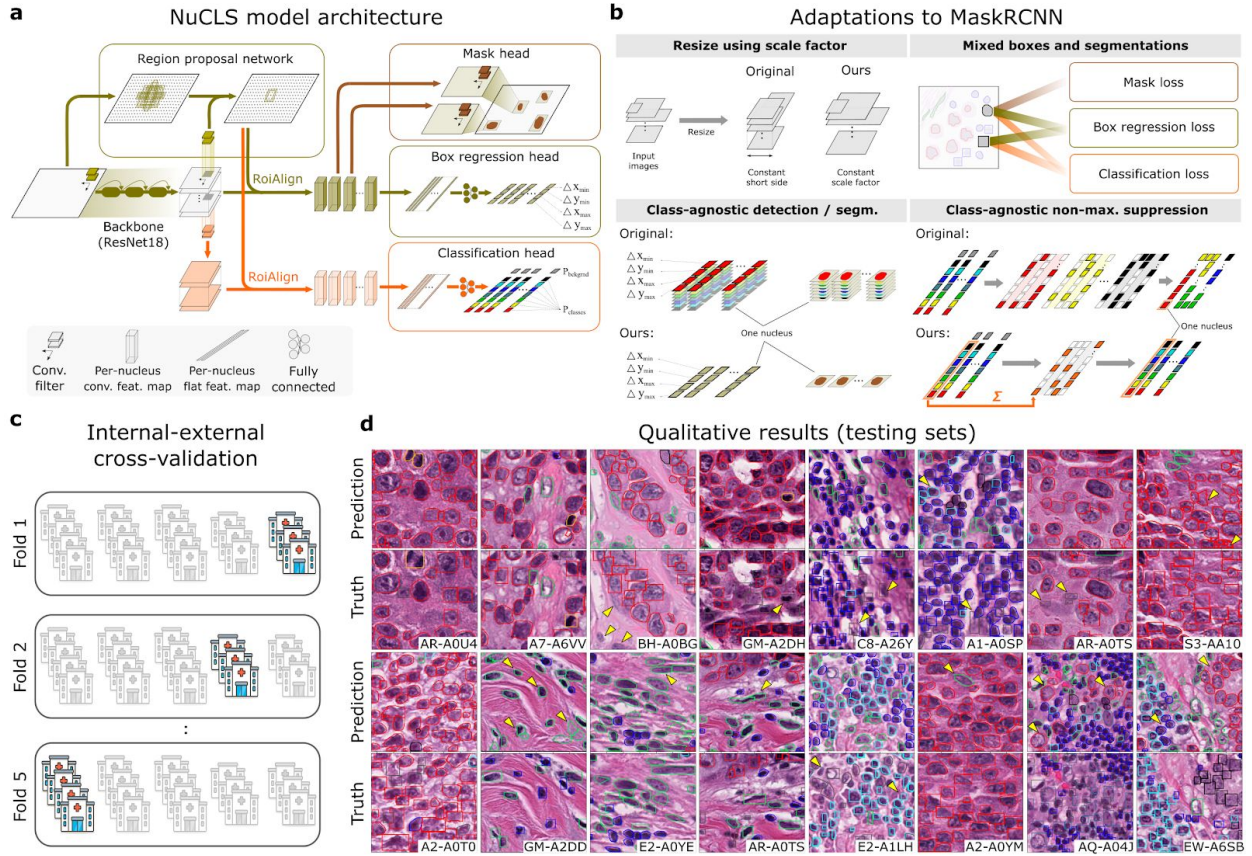

**Figure 6. NuCLS model training and qualitative results. a.** The MaskRCNN architecture was adapted for nucleus detection and classification, allowing some independence of the terminal classification and detection tasks. **b.** Other adaptations we made include: 1. supporting variable-size images at inference while preserving scale and aspect ratio; 2. supporting hybrid training data that mixes bounding boxes and segmentations; 3. simplifying object detection; 4. generating full class probability vectors for each nucleus at inference time (significantly improves performance). **c.** Internal-external cross-validation procedure. The TCGA dataset originates from multiple institutions, and we used this fact to obtain an estimate of the external analytic validity of our models. Fold 1 was used for tuning hyper parameters, while folds 4-5 were used as external testing sets. **d.** Qualitative results from testing sets. Here the truth shown was obtained from the corrected single-rater dataset. Not all discrepancies are algorithm errors, including: i. adjacent nuclei that could conceivably be viewed as a single nucleus; ii. missing annotations, iii. morphologically ambiguous nuclei. Some errors arise from the lack of incorporation of contextual information in our model. Without low power context, macrophages and normal ductal cells may look morphologically similar to tumor cells at high power.

The performance of NuCLS models was consistent with limitations of the training data. Accuracy was lower for classes with higher interrater variability (e.g. plasma cells) or for classes where NPs were not reliable annotators (mitotic figures and macrophages) (Figure S11, Figure 7b,g). Interestingly, we found that super-class accuracy was higher when trained on granular classes than on super-classes (config 2 vs. 6 in Table S5). This indicates that uncommon classes, while noisy, provide signal to improve the function approximation by placing nuclei that look morphologically different (e.g. inactive lymphocytes vs. plasma cells) into different "buckets". We also found that NuCLS models outperform approaches that decouple detection and classification into independent, sequential stages (config 2 vs. 4 in Table S5) [44].

NuCLS models trained on the uncorrected single-rater dataset had slightly lower detection accuracy (AP=72.9±1.1) (Table S6, Figure S12). This suggests that vetting poor detections through correction or exclusion was important; some of the prediction errors could be traced back to a handful of NPs who had outlying detection performance (Figure S13).

**Table 1. Generalization accuracy of the trained NuCLS models.** Models were trained on the corrected single-rater dataset, and evaluated on testing set slides using internal-external cross-validation. Fold 1 acted as the validation set for hyperparameter tuning, and did not contribute to the overall accuracy mean and standard deviation results in the bottom rows. Two types of annotations were used as the testing set: corrected single-rater dataset (top half) and inferred P-truth on the Evaluation dataset (bottom half). Note that the number of testing set nuclei varied by fold because the split happens at the level of hospitals and not nuclei. There were no testing set slides with available inferred P-truth to assess the performance on fold 2. Notice that the classification accuracy is consistently higher when the assessment was done at the level of super-classes. The accuracy of inferred NP-labels can be used as a rough benchmark for comparison. Abbreviations: AP@.5, average precision when a threshold of 0.5 is used for considering a detection to be true; mAP@.5:.95, mean average precision at a range of detection thresholds between 0.5 and 0.95.

| Fold | Detection | | | Segmentation | | | Classification | | | | | |
|---|---|---|---|---|---|---|---|---|---|---|---|---|
| | N | AP @.5 | mAP @.5:.95 | N | Median IOU | Median DICE | N | Super-classes? | Accuracy | MCC | AUROC (micro) | AUROC (macro) |
| **Training: Corrected single-rater dataset; Testing: Corrected single-rater dataset** | | | | | | | | | | | | |
| 1 (Val.) | 6102 | 75.3 | 34.4 | 1389 | 78.5 | 87.9 | 5351 | No | 71.0 | 58.1 | 93.3 | 84.6 |
| | | | | | | | | Yes | 77.5 | 65.2 | 93.7 | 89.0 |
| 2 | 15442 | 74.9 | 33.2 | 3474 | 78.0 | 87.6 | 13597 | No | 70.1 | 56.9 | 93.8 | 83.6 |
| | | | | | | | | Yes | 79.4 | 68.2 | 94.6 | 86.5 |
| 3 | 12672 | 74.0 | 33.8 | 1681 | 80.2 | 89.0 | 11176 | No | 68.6 | 57.0 | 93.5 | 87.1 |
| | | | | | | | | Yes | 79.0 | 68.1 | 94.4 | 89.4 |
| 4 | 8260 | 75.3 | 33.5 | 1948 | 80.9 | 89.5 | 7288 | No | 73.1 | 61.8 | 94.5 | 85.0 |
| | | | | | | | | Yes | 83.9 | 73.5 | 96.1 | 87.4 |
| 5 | 7295 | 74.9 | 31.5 | 1306 | 78.1 | 87.7 | 6294 | No | 61.7 | 47.0 | 89.3 | 79.2 |
| | | | | | | | | Yes | 68.4 | 52.4 | 89.0 | 80.8 |
| Mean (Std) | - | 74.8 (0.5) | 33.0 (0.9) | - | 79.3 (1.3) | 88.5 (0.8) | - | No | 68.4 (4.2) | 55.7 (5.4) | 92.8 (2.0) | 83.7 (2.9) |
| | | | | | | | | Yes | 77.7 (5.7) | 65.6 (7.9) | 93.5 (2.7) | 86.0 (3.2) |
| **Training: Corrected single-rater dataset; Testing: Inferred P-truth (Evaluation dataset)** | | | | | | | | | | | | |
| 1 (Val.) | 209 | 62.9 | 21.0 | 42 | 67.6 | 80.7 | 173 | No | 70.5 | 63.6 | 94.2 | 85.6 |
| | | | | | | | | Yes | 86.1 | 79.0 | 95.7 | 95.6 |
| 3 | 66 | 65.2 | 29.0 | 7 | 76.9 | 86.9 | 52 | No | 63.5 | 42.4 | 80.7 | 85.5 |
| | | | | | | | | Yes | 61.5 | 42.5 | 75.1 | 84.7 |
| 4 | 317 | 71.5 | 32.6 | 82 | 76.2 | 86.5 | 278 | No | 68.0 | 54.3 | 94.3 | 89.3 |
| | | | | | | | | Yes | 84.9 | 75.5 | 96.9 | 92.0 |
| 5 | 213 | 58.3 | 22.9 | 49 | 71.8 | 83.6 | 174 | No | 67.8 | 55.8 | 92.2 | 90.4 |
| | | | | | | | | Yes | 75.3 | 65.6 | 91.4 | 95.2 |
| Mean (Std) | - | 65.0 (5.4) | 28.2 (4.0) | - | 74.9 (2.3) | 85.7 (1.5) | - | No | 66.4 (2.1) | 50.8 (6.0) | 89.1 (6.0) | 88.4 (2.1) |
| | | | | | | | | Yes | 73.9 (9.6) | 61.2 (13.8) | 87.8 (9.2) | 90.6 (4.4) |

## Decision Tree Approximation of Learned Embeddings

From a clinical perspective, nucleus detection and classification are arguably more relevant than precise segmentation of nuclei. Segmentation, however, enables the extraction of quantitative and interpretable morphologic nuclear features, which may contain latent prognostic information and help to discover novel biological associations [5,9–11]. Here we show how segmentation can also be used to enhance the interpretability of nucleus classification models, thereby improving confidence in model decisions, a key requirement for clinical adoption [4].

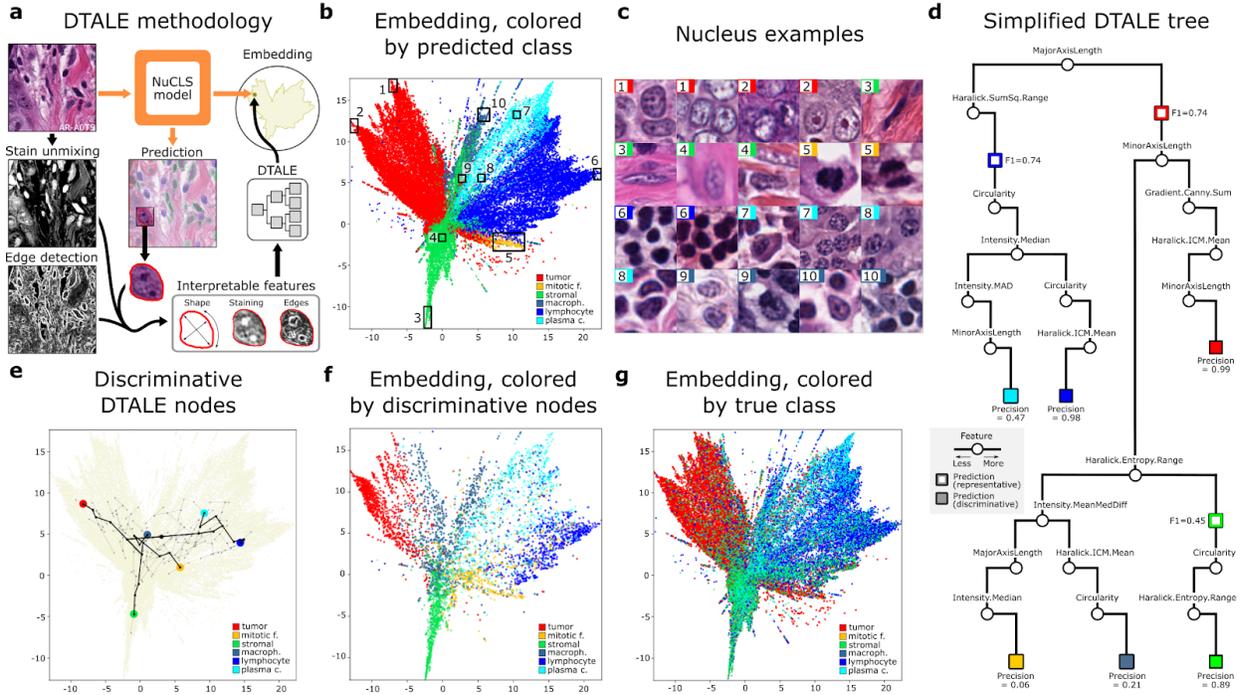

**Figure 7. Explaining NuCLS model decisions using Decision Tree Approximation of Learned Embeddings (DTALE). a.** Illustrative explanation of the DTALE method. Two dimensional UMAP embeddings were obtained from the flattened nucleus classification feature maps. A regression decision tree was then fitted to produce predictions in the embedding space using interpretable nucleus features as inputs. **b.** Classification embeddings, colored by the prediction that the NuCLS model eventually assigns to nuclei. **c.** Sample nuclei from the embeddings in b. Peripheral regions (1-3,5-7,10) contain textbook example nuclei, while nuclei closer to the class boundaries have a more ambiguous morphology. **d.** A simplified version of the DTALE tree, showing representative nodes for the three common classes and discriminative nodes for all classes. To reach a discriminative node, DTALE naturally incorporates more features downstream of the representative nodes. **e.** An overlay of the fitted DTALE tree (light gray) on top of the NuCLS classification embeddings. In black, we show paths to the nodes that allow discriminative, high-precision, approximation of NuCLS decisions. **f.** Nuclei within the embedding, belonging to, and colored by, discriminative DTALE nodes. **g.** Embeddings are colored by the true class. The three super-classes are well-separated, but there is considerable noise associated with the less common classes.

We developed a method called Decision Tree Approximation of Learned Embeddings (DTALE) to interpret black box classification models like NuCLS. DTALE uses segmentation boundaries predicted by NuCLS to extract interpretable features of nuclear morphometry (shape, staining, edges, etc.), that are used to create a decision tree approximation of our black box model (Figure 7). DTALE has an important advantage over existing methods like GRADCAM or LIME in that it provides both an overall explanation of the model decision-making process, as well as explanations for individual nuclei (Figure S14) [46,47]. The outputs of the DTALE tree and the black box model can also be quantitatively compared to evaluate the fidelity of the approximation. We made a distinction between representative explanations of model decisions (e.g. *what features describe most nuclei predicted as tumor?*) and discriminative explanations (e.g. *what features are most specific to tumor predictions?*). The former optimizes for the F1 score, while the latter optimizes for precision (Figure S15).

DTALE fitting accurately explained NuCLS decisions for the most common classes (precision=0.99 (tumor), 0.89 (stroma), 0.98 (sTILs)). The DTALE tree suggests that tumor nuclei are identified by their large size, globular shape, and sharp chromatin edges (i.e. nucleoli or chromatin clumping), that stromal nuclei are identified by their slender shape and rough texture, and that lymphocytes are identified by their small size, circular shape, and hyperchromatic staining. Approximations for less common classes were not reliable, likely due to the NuCLS model relying on visual characteristics that are not reliably captured by our interpretable features [48].

## Outlook

In summary, we have described a scalable crowdsourcing approach that benefits from the participation of NPs to reduce pathologist effort and enables minimal-effort collection of segmentation boundaries. We systematically examined aspects related to interrater agreement and truth inference. We showed that widely used object detection models can be adapted to better fit the nucleus detection paradigm and described a novel technique, DTALE, for enhancing the interpretability and clinical adoption potential of nucleus classification models.

There are important limitations and opportunities to improve on our work. Our results suggest that the participation of NPs can help address the scarcity of pathologists' availability, especially for repetitive annotation tasks. This benefit, however, is restricted to annotating predominant and visually distinctive patterns. Naturally, pathologist input — and possibly full-scale annotation effort- would be needed to supplement uncommon and difficult classes that require greater expertise. We chose to engage medical students and graduates with the presumption that some familiarity with basic histology would help acquire higher quality data. Whether this presumption was warranted, or whether it was possible to engage a broader pool of participants was not investigated. On a related note, while we observed differences based on pathologist expertise, this was not our focus, and we expect to address related questions such as the value of fellowship specialization in future work. Also, we did not measure the time it took participants to create annotations; we relied on the safe assumption that certain types of annotation evidently take less time and effort than others.

We focused our annotation efforts on nucleus detection, as opposed to whole cells. Nuclei have distinct staining (hematoxylin) and boundaries, potentially reducing the interrater variability associated with detection of cell boundaries. This, of course, meant that while data collection was more standardized, modeling was more difficult for some classes. Plasma cells, for instance, are distinguishable not only by their (often non-specific) *cartwheel* nuclei, but also their peri-nuclear halo and abundant cytoplasm. Additionally, our NuCLS modelling did not incorporate low-magnification, region-level patterns. We proposed potential region-cell integration strategies in the past, and we expect this to provide improvements in nuclear classification performance [40].

Finally, we would point out that dataset curation is context-dependent, and is likely to differ depending on the biological problem. Nevertheless, we trust that most of our conclusions have broad implications for other histopathology annotation efforts.

## Methods

### Data sources

The scanned diagnostic slides we used were generated by the TCGA Research Network (https://www.cancer.gov/tcga). They were obtained from 125 patients with breast cancer (one slide per patient). Specifically, we chose to focus on all carcinoma of unspecified type cases that were triple negative (TNBC). The designation of histologic and genomic subtype was based on public TCGA clinical records [24]. All slides were stained with Hematoxylin & Eosin (H&E) and were formalin-fixed and paraffin-embedded (FFPE). The scanned slides were accessed using the Digital Slide Archive repository [35].

Region annotations were obtained from a previous crowdsourcing study that we conducted [24]. Regions of Interest (ROIs), ~1 mm² in size, were assigned to participants by level of difficulty. All region annotations were corrected and approved by a practicing pathologist. These region annotations were used to obtain nucleus class suggestions as described below. Region classes included tumor, stroma, lymphocytic infiltrate, plasmacytic infiltrate, necrosis/debris, and other uncommon regions.

**Algorithmic suggestions**

The process for generating algorithmic suggestions is summarized in Figure S2, and involved the following steps:

**Heuristic nucleus segmentation.** We used simple image processing heuristics to obtain noisy nucleus segmentations [27]. Images were analyzed at scan magnification (40x) with the following steps: 1. Hematoxylin stain unmixing using the Macenko method [49]. 2. Gaussian smoothing followed by global Otsu thresholding to identify foreground nuclei pixels [50]. This step was done for each region class separately to increase robustness. We used a variance of 2 pixels for lymphocyte-rich regions and 5 pixels for other regions. 3. Connected-component analysis was used to split the nuclei pixel mask using 8-connectivity and a 3x3 structuring element [51]. 4. We computed the euclidean distance from every nucleus pixel to the nearest background pixel, and found the peak local maxima using a minimum distance of 10 [52]. 5. A watershed segmentation algorithm was used to further split the connected components from step 3 into individual nuclei using the local maxima from step 4 as *markers* [53,54]. 6. Any object < 300 pixels in area was removed.

**Bootstrapping noisy training data.** Region annotations were used to assign a noisy class to each segmented nucleus. This was based on the observation that although tissue regions usually contain multiple cell types, there is often a single predominant cell type: tumor regions / tumor cells, stromal regions / fibroblasts, lymphocytic infiltrate / lymphocytes, plasmacytic infiltrate / plasma cells, other regions / other cells. One exception to this direct mapping are stromal regions, which contain a large number of sTILs in addition to fibroblasts. Within stromal regions a nucleus was considered a fibroblast if it had a spindle-like shape with aspect ratio between 0.4 and 0.55 and circularity between 0.7 and 0.8.

**MaskRCNN refinement of bootstrapped suggestions.** A MaskRCNN model with a Resnet50 backbone was used as a function approximator to refine the bootstrapped nucleus suggestions. This model was trained using randomly cropped 128x128 tiles where the number of nuclei was limited to 30.

**FOV sampling procedure.** Regions-of-interest were tiled into non-overlapping potential FOVs (see *Data sources*). These were selected for inclusion in our study based on a predefined stratified sampling criteria. 16.7% of FOVs were sampled such that the majority of refined suggestions were a single class, e.g. almost all suggestions are tumor. 16.7% were sampled to favor FOVs with two almost equally-represented classes, e.g. a large number of tumor *and* fibroblast suggestions. 16.7% of FOVs were sampled to favor discordance between the bootstrapped suggestions and MaskRCNN-refined suggestions, e.g. stromal region with sTILs.

The remaining 50% of FOVs were randomly sampled from the following pool, with the intent of favoring the annotation of difficult nuclei: a) the bottom 5% of FOVs containing high numbers of nuclei with low MaskRCNN confidence; b) and the top 5% of FOVs containing extreme size detections, presumably clumped nuclei.

### Annotation procedure and data management

The annotation protocol used is provided in the supplement. We asked the participants to annotate the single-rater dataset first because this also acted as their de-facto training. Participants were blinded to the multi-rater dataset name to avoid biasing them. The Unbiased control was annotated first for the same reason. A summary of the data management procedure is provided below.

**HistomicsUI.** We used the Digital Slide Archive, a web-based data management tool, to assign slides and annotation tasks (digitalslidearchive.github.io) [35]. HistomicsUI, the associated annotation interface, was used for creating, correcting, and reviewing annotations. Using a centralized setup avoids participants installing software and simplifies dissemination of images, control over view/edit permissions, monitoring of progress, and collection of results. The annotation process is illustrated in the following video: https://youtu.be/HTvLMyKYyGs. The process of pathologist review of annotations is illustrated in Figure S1.

**HistomicsTK API.** The HistomicsTK Restful Application Programming Interface (API) was used to programmatically manage data, users, and annotations. This includes uploading algorithmic suggestions, downloading participant annotations, and scalable correction of systematic annotation errors where appropriate.

### Obtaining labels from multi-rater datasets

**Obtaining anchor proposals.** We implemented a constrained agglomerative hierarchical clustering process to obtain anchor proposals (Figure 2a). The algorithm is summarized in Figure S16. In order to have a single frame of reference for comparison, annotations from all participants and for all multi-rater datasets were clustered. After clustering, we used two rules to decide which anchor proposals corresponded to actual nuclei (for each multi-rater dataset independently): 1. A nucleus must be detected by at least two pathologists. 2. The inferred P-truth must concur that the anchor is a nucleus.

**Inference of NP-labels & P-truth.** We used the Expectation-Maximization (EM) framework described by Dawid and Skene, and implemented in Python by Zheng *et al* [36,37,55]. Each participant was assigned an initial quality score of 0.7, and 70 EM iterations were performed. As illustrated in Figure 2b, *undetected* was considered to be a nucleus class for the purposes of P-truth/NP-label inference. The same process was used to infer whether the boundary of an algorithmic suggestion was accurate. In effect, the segmentation accuracy was modeled as a binary variable (clicked vs not clicked), and the EM procedure was applied to infer its value.

## Class grouping

We defined two levels of grouping for nuclei classes as illustrated in Figure 2c. This was done for both the single-rater and multi-rater dataset annotations. When handling probabilities, the group probability was calculated by summing probabilities across group members. This applied to grouping classes from the EM inference algorithms, as well as grouping NuCLS model prediction probabilities to obtain super-class level classification predictions.

## Participant agreement

Overall interrater agreement was measured using Krippendorph's alpha statistic, implemented in Python by Santiago Castro and Thomas Grill [56–58]. This statistic was chosen because of its ability to handle missing values [59]. Pairwise interrater agreement was measured using Cohen's Kappa statistic [60]. Likewise, self-agreement was measured using Cohen's Kappa. All of these measures range from -1 (perfect disagreement) to +1 (perfect agreement). A kappa (or alpha) value of zero represents agreement that is expected by random chance. We used thresholds set by Fleiss for defining slight, fair, moderate, substantial, and near perfect agreement [59].

## Annotation redundancy simulations

We performed simulations to measure the impact of the number of NPs assigned to each FOV on the accuracy of NP-label inference (Figure 3e). We kept the total number of NPs constant at 18 and randomly removed annotations to obtain a desired number of NPs per FOV. No constraints were placed on how many FOVs any single NP had. This simulated the realistic scenario where participants are allowed to annotate as many FOVs as they want, and our decision-making focuses on FOV assignment. For each random realization, we calculated the inferred NP-labels using EM and measured accuracy against the static P-truth. This process was repeated for 1000 random realizations per configuration.

## NuCLS model

Our NuCLS model modifies the Pytorch implementation of the MaskRCNN architecture [39,61]. Hyperparameters are provided in the Table S7.

**Resize using scale factor.** MaskRCNN resizes input images to have a constant short side. While this may work for datasets where the variability in image size is modest, or where the camera distance is variable, it is not suitable in computational pathology applications where large tile sizes are favorable for efficient and scalable inference. Resizing to a constant short side would shrink nuclei during inference. To remedy this NuCLS resizes using a *scale factor*, instead, thus preserving the nuclear size and aspect ratio at inference for any tile size.

We used a scale factor of 4.0, meaning that images were digitally zoomed to a 0.05 micron-per-pixel resolution before being analyzed. This corresponded to a sTILs diameter of ~4.4 "pixels" in the feature map generated by the ResNet18 backbone. As a form of scale augmentation, we jittered this scale factor by up to 10% during training.

**Training with hybrid datasets.** Our annotation protocol generates a mixture of manually placed bounding boxes and approved suggestions of segmented nuclei. We train from this data by ignoring bounding boxes when calculating the mask loss.

**Specialized classification convolutions.** Four extra convolutional filters were applied to the feature map output from the ResNet18 backbone [62]. The filters had a kernel size of 3, a stride of 1, and a dilation and padding of 1 to preserve feature map size ([Figure 6a](#)).The resultant feature map was *only* used for classification and only contributed to the classification loss. The same procedure used for box regression was used for classification: 1. ROIAlign to obtain per-object convolutional feature maps; 2. flattening of the feature map; 3. passage through a single fully-connected layer.

**Class-agnostic detection & segmentation.** Both the box regression output and nucleus masks were simplified and made classification-agnostic. We relied on the fact that nucleus shapes and sizes are fairly homogeneous to simplify the learning problem and preserve classification probability vectors at inference. Specifically, we relied on a global non-maximum suppression (NMS) process ([Figure 6b](#)). We summed the classification probabilities for all classes (i.e. everything *except* background), and concatenated all these "objectness" scores for each FOV. An NMS process was then carried out as usual. That is, boxes were sorted by objectness score, and if a box overlapped with a higher-scoring box by more than a particular IOU threshold (0.2 in our case), it was removed.

**Data augmentation.** Previous research has shown that the combined use of color normalization and augmentation improves performance of deep learning models in histopathology applications [63]. All FOVs were color normalized using the Macenko method before training began [49]. During training, FOVs also underwent a stain augmentation routine [64]. This augmentation routine randomly perturbed the hematoxylin and eosin channels each time the image was loaded, using a sigma of 0.5 for the random uniform distribution. The HistomicsTK package was used for both the color normalization and augmentation operations ([digitalslidearchive.github.io](#)). Additionally, a random 300x300 pixel region was cropped on-the-fly during training.

**Handling class imbalance.** Nucleus class imbalance was mitigated by weighted random sampling with replacement. With the exception of ambiguous nuclei, which received zero weight, class weights were inversely proportional to the frequency of occurrence in the training set. Since we load data on a per-FOV basis, each FOV $f$ was assigned a sampling weight $W_f$ that favors FOVs with a high density of uncommon nuclear classes, as follows:

$$W_f = U_f \ / \ \sum_{i=1}^{F} U_i \ \text{, where} \ U_f = \sum_{c=1}^{C} \left( W_c N_{cf} \right) \ / \ A_f$$

$$W_c = V_c \ / \ \sum_{i=1}^{C} V_i \ \text{, where} \ V_c = 1 \ / \ \sum_{f=1}^{F} N_{cf}$$

Where $C$ is the number of classes, $F$ is the number of FOVs in the training set, $W_c$ is the weight assigned to class $c$, $N_{cf}$ is the number of nuclei of class $c$ in FOV $f$, and $A_f$ is the area of FOV $f$.

**Internal-external cross-validation.** Training and testing data were separated at the level of hospitals/institutions, as described in the discussion section and in Figure 6c. Institutions having <9 patients from our patient cohort were considered to be small institutions.

**Matching detections.** To determine the segmentation and classification accuracy, algorithmic detections were matched to ground truth using linear sum assignment from the Scipy library [65].

**Decision Tree Approximation of Learned Embeddings**

DTALE relies on the fact that MaskRCNN (and by extension, our NuCLS model) learns to predict object segmentation boundaries as well as their classifications [39]. The DTALE procedure involves four steps (Figure 7a): 1. learning embeddings 2. generating interpretable features 3. fitting the decision tree and 4. calculating node statistics.

**Learned embeddings.** Starting with a trained NuCLS model, we extracted the terminal, per-nucleus, 1024-dimensional classification feature vectors (just before the logits). Hyperbolic UMAP was applied to these features to generate a two-dimensional (2D) embedding [66].

**Interpretable features.** The same FOVs that were input into the NuCLS model were processed to enable extraction of interpretable features. Macenko stain unmixing was used to separate the hematoxylin channel [49]. Both the hematoxylin intensity channel and the segmentation mask predictions from the NuCLS model were input into the HistomicsTK function *compute_nuclei_features*, which uses image processing operations to extract feature vectors encoding 62 morphologic features describing shape, intensity, edges, and texture (Table S8) [67–69].

**Regression decision tree.** A regression decision tree was fitted to produce predictions in the embedding space using the interpretable features as inputs [70]. This maps the interpretable features directly into the 2D embedding space to connect morphology with NuCLS model behavior. The rationale for using a regression tree, as opposed to a classification tree, is two-fold. First, any accurate classification model will produce similar classification decisions. In contrast, the 2D embedding is a compressed version of a 1024-feature space that is highly specific to our trained NuCLS model. Second, using a regression tree allows us to produce fine-grained *within-class* explanations for individual nuclei (see [Figure S14](#)). This technique is broadly similar to some existing works that use soft decision trees to approximate deep-learning model behavior [71]. We constrained the decision tree to a maximum depth of 7 and a minimum of 250 nuclei per leaf.

**Node fit statistics.** Once the DTALE tree was fitted, we traversed nodes to find paths that best represented NuCLS class predictions. For each classification class $C_j$, and for each tree node $N_i$, we calculate precision, recall, and F1 scores for the downstream subtree as if all nuclei were classified as $C_j$ and using actual NuCLS model classifications as ground truth. This generates an F1 and precision score for each node/class pair. For each class we identify the node with the highest F1 score as the most representative of NuCLS model predictions for that class, whereas the node with the highest precision corresponds to interpretable features that are the most discriminative for that class.

## Software

Data management, machine learning models and plotting were all implemented using Python 3+. Pytorch and Tensorflow libraries were used for various deep-learning experiments. Scikit-learn, Scikit-image, OpenCV, HistomicsTK, Scipy, Numpy and Pandas libraries were used for matrix and image processing operations. Openslide library and HistomicsTK API were used for interaction with whole-slide images.

## Statistical tests

The Mann-Whitney U test was used for unpaired comparisons. The Wilcoxon signed-rank test was used for paired comparisons. Confidence bounds for ROC curves were obtained by bootstrap sampling [72,73].

# Acknowledgements

This work was supported by the U.S. National Institutes of Health National Cancer Institute grants U01CA220401 and U24CA19436201. We would like to acknowledge with gratitude the contributions made by the following participants: Eman Elsayed Sakr (El-Matariya Teaching Hospital, Egypt), Joumana Ahmed (Cairo University, Egypt); Mohamed Zalabia and Ahmed S. Badr (Menoufia University, Egypt); Ahmed M. Afifi (Ain Shams University, Egypt); Esraa B. Ghadban (Damascus University, Syria); Mahmoud A. Hashim (Baylor College of Medicine, USA). We are thankful to Uday Kurkure, Jim Martin, Raghavan Venugopal, Joachim Schmidt (Roche Tissue Diagnostics, USA) and Michael Barnes (Roche Diagnostic Information Solutions, USA) for support and discussions. We also thank Brian Finkelman for constructive feedback on the interrater analysis. Finally, we acknowledge with gratitude Jeff Goldstein and other members of the Cooper research group at Northwestern for constructive feedback and discussion.

## Author contributions

M.A. and L.A.D.C. conceived the hypothesis, designed the experiments, performed the analysis, and wrote the manuscript. D.M. and D.A.G. contributed support for the Digital Slide Archive software and database. P.M. contributed to the analysis of DTALE embeddings. B.D. and D.J. provided ideas for the interrater analysis. M.A. and M.A.T.E. were the study coordinators and corrected the single-rater dataset. H.E. provided feedback and approved the corrected single-rater dataset. E.H. provided manual nucleus segmentation data. H.E., H.H., and E.H. are senior pathologists and provided multi-rater annotations. L.A.A., K.H.M., P.A.P., and L.E.H. are junior pathologists and provided multi-rater annotations. M.A.T.E., A.M.A., M.A.A., A.M.E., R.A.S., A.R., A.M.S., A.M.A., I.A.R., A.A., N.M.E., A.A., A.F., A.E., A.G.E., Y.A., Y.A.A., A.M.R., M.K.N., M.A.T.E., A.A., A.G., and M.E. are non-pathologists and provided single- and multi-rater annotations. All experience designations are based on the time of annotation. All authors reviewed the manuscript draft.

## Conflicts of interest

None to disclose.

## Code availability

Relevant code can be found at github.com/CancerDataScience/NuCLS.

# Supplementary tables

**Table S1. Definitions and abbreviations used.** The following paper from the Digital Pathology Association can be consulted for an expanded list of relevant concepts: *Abels, E. et al, The Journal of Pathology. 2019. 249: 286–294*.

| Term | Abbr. | Definition |
|------|-------|------------|
| **Basic definitions** | | |
| Whole slide image | WSI | High-resolution scanned image of a histopathology slide. Most WSIs of solid tumors are scanned at a 20-40x magnification and are extremely large (~80k pixels side) |
| Annotation | - | Manual markup of image to indicate the location, boundary, or class of an anatomical structure. Examples include a point at the centroid of a nucleus, a bounding box that indicates the extent of a nucleus, or tracing of the nucleus boundary. |
| Segmentation | - | A boundary delineating the edge of a structure like a histologic region or a nucleus. |
| Ground truth | - | The true location/boundary/class of a particular nucleus. This term is used loosely in this paper to refer to the truth against which the deep-learning models are evaluated. This truth will be different under different circumstances, depending on the experiment being discussed. |
| Region of interest | ROI | A ~1 mm$^2$ region of a WSI that FOVs are selected from. Each ROI is accompanied by low power annotations of tissue regions used for generating suggestions. |
| Field of view | FOV | A ~65 x 65 µm field selected from within an ROI. FOVs were annotated at high-power to indicate the location and class of all nuclei contained in the FOV. |
| Application Programming Interface | API | A set of functions that allows developers to programmatically interact with a database or other software. |
| **Participant groups** | | |
| Non-pathologists | NPs | Medical students/graduates who did not receive pathology residency training. |
| Junior pathologists | JPs | Pathology residents with < 2 years of anatomical pathology training. |
| Senior pathologists | SPs | Attendings or pathology residents with > 2 years of anatomical pathology training. |
| Pathologists | Ps | Junior or senior pathologists. |
| **Datasets** | | |
| Hybrid dataset | - | A dataset where participants click accurate segmentation boundary suggestions, and draw bounding boxes around all other nuclei. The resultant dataset contains a mixture of segmentation boundaries and bounding boxes. |
| Single-rater dataset | - | A collection of FOVs that were annotated by NPs in a single-rater manner. NPs received pathologist feedback during the annotation process. NPs were shown both region (low-power) and nucleus (high-power) suggestions while annotating. |
| Corrected single-rater dataset | - | A subset of single-rater dataset FOVs (approximately half) whose annotations have been manually corrected by study coordinators based on feedback from a senior pathologist. All corrected single-rater dataset annotations were approved by a senior pathologist. |
| Uncorrected single-rater dataset | - | Single-rater dataset FOVs whose annotations were not manually corrected. The quality of these annotations is participant-dependent. |
| Multi-rater datasets | - | A collection of FOVs that were annotated by multiple participants under different experimental conditions. NPs were not given feedback on these FOVs. These are used for interrater comparisons. |

| Evaluation dataset | - | A multi-rater dataset where MaskRCNN refined algorithmic suggestions were shown to the participants. Refinement was applied to bootstrap suggestions to improve quality. These suggestions were the same type used in single-rater dataset annotation. |
|---|---|---|
| Bootstrap control | - | A multi-rater dataset where noisy bootstrapped algorithmic suggestions were shown to the participants. These suggestions were generated using a heuristic segmentation algorithm and processing of low power region annotations and shape data from segmentation. |
| Unbiased control | - | A multi-rater dataset where no annotation suggestions were shown to the participants. This was the first multi-rater dataset annotated to obtain annotations thare were not biased by algorithmic suggestions. |
| **Nucleus suggestions and labels** | | |
| Bootstrapped suggestions | - | A set of noisy nuclear boundary suggestions using simple image processing heuristics. Each boundary also had an associated classification suggestion, inherited from the histologic region in which the presumed-nucleus resides. For example, a suggested boundary in a tumor region would be associated with a tumor classification suggestion. These were an intermediate step in the production of refined suggestions (see below), and were only shown to participants for the Bootstrap control dataset. |
| MaskRCNN refined suggestions | - | The result of fitting a MaskRCNN model to the bootstrap suggestion. MaskRCNN acts as a function approximator to smooth out noise. These were shown to participants for the single-rater and Evaluation datasets. |
| Label | - | This term is used in the broad sense, as in *labeled data* used for supervised machine learning. A label is a tag associated with a potential nucleus location (anchor proposal, defined below). Labels include assessment of whether an anchor proposal corresponds to a nucleus (i.e. detection), what class to assign (e.g. tumor) and whether or not the suggested segmentation boundary is correct. |
| Anchor proposal | - | A *potential* bounding-box location of a nucleus. Anchor proposals are generated by clustering annotations from multi-rater datasets. |
| Class | - | A type of label that assigns a nucleus to a set of predefined biological categories (e.g. tumor, fibroblast, etc). |
| Raw nucleus classes | - | The set of 12 nucleus classes that were directly obtained from the participants, without class grouping. |
| Nucleus classes | - | A set of 7 nucleus classes, obtained by grouping related raw classes together. |
| Nucleus super-classes | - | Three clinically-salient nucleus classes (tumor, stroma, sTILs), obtained by grouping nucleus classes. |
| Uncommon nucleus classes | - | Any raw nucleus classes other than tumor, fibroblasts, and lymphocytes. |
| Inferred pathologist truth | P-truth | A single label that is generated from the analysis of multi-rater datasets using pathologist annotations. For each anchor proposal from clustering we use EM to infer whether the proposal is an actual nucleus, the class, and the correctness of the suggested boundary. This was used to measure the accuracy of NP annotations and NP-label. |
| Inferred non-pathologist label | NP-label | A single label that is generated from the analysis of multi-rater datasets using NP annotations (see inferred P-truth for comparison). |
| **Machine learning and image processing** | | |
| Convolutional neural network | CNN | A deep-learning model that operates on image data. |
| MaskRCNN | - | A CNN model that learns to jointly predict nucleus bounding-box localization, segmentation, and class. |
| Nucleus Classification, Localization and Segmentation | NuCLS | A modified MaskRCNN model that uses hybrid datasets from our proposed crowdsourcing approach to predict nucleus bounding box locations, segmentation boundaries, and classifications. |
| Agglomerative hierarchical clustering | - | A bottom-up clustering approach that builds a hierarchy of clusters starting with each data point as its own cluster, and grouping data points and clusters by similarity. |
| Expectation-Maximization | EM | An iterative method for estimating the parameters of a statistical model by maximizing a *likelihood* measure. Used to simultaneously estimate participant reliability and nucleus locations, class, and correctness of boundaries. |
| Heuristic nucleus segmentation | - | Delineation of nuclear boundaries using simple image processing operations that have no dependence on annotation data (unlike machine learning models). This was used to generate bootstrapped algorithmic suggestions or segmentation boundaries. |
| Interpretable features | - | A set of engineered features that use image processing operations to describe nuclei. These include interpretable (i.e. easy to explain) morphological descriptors of shape, texture and edges. |
| Decision tree | - | A machine learning model that learns a sequence of discrete decisions to make predictions. Regression decision trees learn discrete thresholds at each node (e.g. *Is the major axis greater than x µm?*). |
| Decision tree approximation of learned embeddings | DTALE | A technique we devised for increasing the transparency of black box deep learning models. |
| **Measures of accuracy and agreement** | | |
| Intersection over union | IOU | Quantitative measure of overlap of prediction and truth. |
| DICE coefficient | - | Similar to IOU, it is a measure of overlap of prediction and truth. |
| Area under Receiver-Operator Characteristic (ROC) curve | AUROC | It is a measure of accuracy, where a value of 0.5 corresponds to random chance and a value of 1.0 is the maximum. There are two ways of obtaining this value: <br> - *Micro-average:* Is the overall accuracy, where different nucleus classes contribute to the result in proportion to their abundance in the dataset <br> - *Macro-average:* Is the class-balanced accuracy, where different nucleus classes are equally weighted, such that an uncommon class like macrophages will have the same contribution as a common class like sTILs. |
| Average precision | AP | Area under precision-recall curve used to measure detection performance. AP@.5 refers to area measured with a minimum IOU of 0.5 for defining correct detections. mAP@.5:.95 is a more stringent measure that averages areas for a range of IOU thresholds from 0.5 to 0.95. |
| F1 score | - | The harmonic mean of precision and recall values. |
| Matthiew's Correlation Coefficient | MCC | A balanced measure of classification accuracy that takes into account all components of the confusion matrix, including true negatives (unlike the F1 score). |
| Cohen's Kappa statistic | - | A measure of agreement between two participants, ranging from -1 (perfect disagreement) to +1 (perfect agreement). |
| Krippendorph's Alpha statistic | - | A multi-rater generalization of Cohen's Kappa, which handles missing values. |

**Table S2. Accuracy of algorithmic suggestions.** The accuracy is measured against the corrected single-rater dataset. MaskRCNN refinement of the bootstrapped algorithmic suggestions results in better detection suggestions. Low-power region-based classification was more accurate than MaskRCNN-derived classes. Note, however, that this was FOV-dependent, and there were some FOVs in which the MaskRCNN prediction was better than relying on low-power regions for classification.

| Stage | Class | | N | Accuracy | MCC | F1 | Precision | Sensitivity | Specificity |
|---|---|---|---|---|---|---|---|---|---|
| Bootstrap suggestions | Detection | | 58598 | **18.8** | - | **31.7** | **40.4** | **26.1** | - |
| | Classification (region-inherited) | **Overall** | 11029 | **86.8** | **77.9** | - | - | - | - |
| | | Tumor | | 95.6 | 91.2 | 95.2 | 93.6 | 96.8 | 94.7 |
| | | Stromal | | 90.3 | 21.5 | 12.4 | 80.0 | 6.7 | 99.8 |
| | | sTILs | | 89.6 | 80.3 | 89.2 | 82.6 | 97.0 | 83.7 |
| | | Other | | 98.2 | 16.8 | 17.7 | 16.9 | 18.6 | 99.0 |
| Suggestions after MaskRCNN refinement | Detection | | 75908 | **31.5** | - | **47.9** | **47.6** | **48.1** | - |
| | Classification (region-inherited) | **Overall** | 23874 | **78.9** | **67.6** | - | - | - | - |
| | | Tumor | | 93.5 | 85.9 | 91.1 | 90.1 | 92.1 | 94.2 |
| | | Stromal | | 82.0 | 46.1 | 57.2 | 53.4 | 61.6 | 87.0 |
| | | sTILs | | 83.6 | 66.4 | 80.2 | 84.2 | 76.6 | 88.9 |
| | | Other | | 99.3 | 30.5 | 25.9 | 56.0 | 16.9 | 99.9 |
| | Classification (MaskRCNN prediction) | **Overall** | | **69.1** | **52.7** | - | - | - | - |
| | | Tumor | | 83.2 | 63.0 | 74.9 | 82.1 | 68.8 | 91.4 |
| | | Stromal | | 82.2 | 26.3 | 17.5 | 91.3 | 9.6 | 99.8 |
| | | sTILs | | 75.5 | 58.1 | 77.4 | 64.5 | 96.8 | 59.0 |
| | | Other | | 97.9 | 12.0 | 11.9 | 8.5 | 19.9 | 98.5 |

**Table S3. NuCLS model tuning for the nucleus detection task on the validation set (fold 1).** After passage through the model backbone, the feature map is markedly smaller than original images due to the max pooling operations. This means that without digital zooming, the diameter of a 'typical' small nucleus, say TILs, is very small in the feature map. As a consequence, when the object-specific part of the feature map is pooled using ROIAlign, there is very little information to use for box regression or classification. Abbreviations: MPP, microns-per-pixel; AP @ 0.5, average precision when a threshold of 0.5 is used for considering a detection to be true.

| Scale factor | Equivalent MPP | Backbone | TILs diameter (image, pixels) | TILs diameter (featmap, 'pixels') | AP @ 0.5 |
|---|---|---|---|---|---|
| 1 | 0.2 | Resnet18 | 30 | 1.1 | 61.7 |
| 1 | 0.2 | Resnet34 | 30 | 1.1 | 63 |
| 1 | 0.2 | Resnet50 | 30 | 1.1 | 62 |
| 2.67 | 0.075 | Resnet18 | 80 | 3 | 76.4 |
| 2.67 | 0.075 | Resnet34 | 80 | 3 | 74.3 |
| 2.67 | 0.075 | Resnet50 | 80 | 3 | Mem.Err. |
| 4 | 0.05 | Resnet18 | 120 | 4.4 | 75 |
| 4 | 0.05 | Resnet34 | 120 | 4.4 | 72.9 |
| 4 | 0.05 | Resnet50 | 120 | 4.4 | Mem.Err. |

**Table S4. Generalization accuracy of the trained NuCLS models - broken down by superclass.**

| Fold | N | MCC | | | | AUROC | | | | |
|---|---|---|---|---|---|---|---|---|---|---|
| | | Overall | Tumor | Stromal | sTILs | Micro-avg. | Macro-avg. | Tumor | Stromal | sTILs |
| Training: Corrected single-rater dataset; Testing: Corrected single-rater dataset | | | | | | | | | | |
| 1 (Val.) | 5351 | 65.2 | 72.9 | 47.1 | 73.7 | 93.7 | 89.0 | 94.2 | 83.2 | 95.3 |
| 2 | 13597 | 68.2 | 73.7 | 53.0 | 76.6 | 94.6 | 86.5 | 94.5 | 87.4 | 96.2 |
| 3 | 11176 | 68.1 | 74.9 | 46.9 | 77.9 | 94.4 | 89.4 | 96.1 | 84.3 | 95.7 |
| 4 | 7288 | 73.5 | 80.6 | 56.9 | 79.6 | 96.1 | 87.4 | 97.2 | 89.1 | 95.9 |
| 5 | 6294 | 52.4 | 57.4 | 40.7 | 60.1 | 89.0 | 80.8 | 88.8 | 80.7 | 91.0 |
| Mean (Std) | - | 65.6 (7.9) | 71.7 (8.6) | 49.4 (6.1) | 73.5 (7.8) | 93.5 (2.7) | 86.0 (3.2) | 94.2 (3.2) | 85.4 (3.2) | 94.7 (2.1) |
| Training: Corrected single-rater dataset; Testing: Inferred P-truth (Evaluation dataset) | | | | | | | | | | |
| 1 (Val.) | 173 | 79.0 | 88.0 | 73.0 | 78.6 | 95.7 | 95.6 | 97.7 | 94.4 | 95.5 |
| 3 | 52 | 42.5 | 38.5 | 26.3 | 73.9 | 75.1 | 84.7 | 87.1 | 83.0 | 90.9 |
| 4 | 278 | 75.5 | 77.8 | 53.1 | 90.2 | 96.9 | 92.0 | 96.4 | 91.9 | 99.2 |
| 5 | 174 | 65.6 | 60.0 | 67.1 | 72.1 | 91.4 | 95.2 | 96.6 | 92.2 | 97.9 |
| Mean (Std) | - | 61.2 (13.8) | 58.8 (16.1) | 48.8 (16.9) | 78.8 (8.2) | 87.8 (9.2) | 90.6 (4.4) | 93.4 (4.4) | 89.0 (4.3) | 96.0 (3.6) |

**Table S5. NuCLS model tuning for the nucleus classification task on the validation set (fold 1).** Empty entries correspond to metrics which were not applicable for the configuration (config) being studied. Classification AUROC statistics were not possible for configs where each nucleus had a single classification as opposed to a classification probability vector. Configs where the model was trained on super-classes do not have accuracy statistics for the main classes. On the other hand, when models were trained on the main classes, super-class predictions were easily obtained by aggregating the predicted class probabilities.

| Config | Detection AP @.5 | Overall classification accuracy | | | | | | Classification accuracy breakdown (AUROC) | | | | | | | | |
|---|---|---|---|---|---|---|---|---|---|---|---|---|---|---|---|---|
| | | MCC | | Micro | | Macro | | Tumor | | | Stromal | | | sTILs | | |
| | | Supercl.? | | Supercl.? | | Supercl.? | | Subclasses | | Supercl ass | Subclasses | | Supercl ass | Subclasses | | Supercl ass |
| | | No | Yes | No | Yes | No | Yes | Non-mit otic | Mitoti c | | Stro mal | Macrop hage | | Lympho cyte | Plasma cell | |
| 1 | 70 | 1.8 | -3 | - | - | - | - | - | - | - | - | - | - | - | - | - |
| 2 | 74.5 | 57 | 65 | 93.4 | **94.3** | **85.2** | **88.2** | 93.1 | **91.5** | 93.2 | 88.8 | 71 | **83.6** | 95 | 78.6 | 95 |
| 3 | **75.4** | **59.6** | **66** | **93.5** | 93.7 | 84.7 | 85.2 | **94.2** | 90.6 | **94.5** | **89.1** | **73.5** | 82 | **95.2** | 84.2 | **95.7** |
| 4 | 72.2 | 52.6 | 60.9 | 91 | 92.3 | 82.4 | 83.6 | 92.5 | 90.8 | 92.1 | 86.7 | 61.7 | 78.9 | 94.7 | 82.9 | 93.4 |
| 4+ | 72.2 | 54.5 | 62.5 | 90.3 | 91.9 | 84.1 | 85.8 | 92.2 | 88.5 | 92 | 88.1 | 68.4 | 81.5 | 93.7 | **84.4** | 93.4 |
| 5 | 72.6 | - | -5 | - | - | - | - | - | - | - | - | - | - | - | - | - |
| 6 | 74.8 | - | 63.6 | - | 93.5 | - | 85.9 | - | - | 92.8 | - | - | 81.3 | - | - | 95 |
| 7 | 72.2 | - | 63.1 | - | 93.1 | - | 82.8 | - | - | 91.9 | - | - | 81 | - | - | 94.9 |
| 7+ | 72.2 | - | 64.8 | - | 92.7 | - | 83.7 | - | - | 93.1 | - | - | 83.1 | - | - | 94.8 |

**Config 1:** Slightly modified MaskRCNN implementation (with mixed boxes and segmentations training)
**Config 2:** Config 1, but with class-agnostic detection and non-maximum suppression
**Config 3:** Config 2, but with 4 extra convolutions that specialize in classification
**Config 4:** Config 1 for nucleus detection, then an independent nucleus classification model using thumbnails of detected nuclei
**Config 4+:** Same model from config 4, but with test-time augmentation (random shift) at the classification stage
**Config 5:** Config 1 but trained using supercategories
**Config 6:** Config 2 but trained using supercategories
**Config 7:** Config 4 but trained using supercategories
**Config 7+:** Same model from config 7, but with test-time augmentation (random shift) at the classification stage

**Table S6. Accuracy of the trained NuCLS models on the uncorrected single-rater dataset.**

| Fold | Detection | | | Segmentation | | | Classification | | | | | |
|------|-----|--------|-------------|-----|---------------|----------------|------|-------------------|----------|------|------------------|------------------|
| | N | AP @.5 | mAP @.5:.95 | N | Median IOU | Median DICE | N | Super-classes ? | Accuracy | MCC | AUROC (micro) | AUROC (macro) |
| **Training: Uncorrected single-rater dataset; Testing: Corrected single-rater dataset** | | | | | | | | | | | | |
| 1 | 6102 | 71.6 | 29.8 | 1371 | 75.9 | 86.3 | 5236 | No | 71.3 | 57.6 | 92.3 | 83.9 |
| | | | | | | | | Yes | 76.3 | 62.6 | 92.9 | 87.3 |
| 2 | 15442 | 73.1 | 31.2 | 3467 | 76.1 | 86.4 | 13517 | No | 62.8 | 49.2 | 91.8 | 77.8 |
| | | | | | | | | Yes | 74.0 | 61.7 | 92.4 | 84.6 |
| 3 | 12672 | 71.1 | 32.0 | 1663 | 80.7 | 89.3 | 11025 | No | 67.2 | 54.9 | 92.5 | 80.5 |
| | | | | | | | | Yes | 78.8 | 66.4 | 93.9 | 85.4 |
| 4 | 8260 | 73.7 | 32.0 | 1953 | 79.2 | 88.4 | 7309 | No | 71.6 | 58.9 | 93.8 | 81.5 |
| | | | | | | | | Yes | 81.6 | 69.1 | 95.2 | 85.4 |
| 5 | 7295 | 73.7 | 32.7 | 1323 | 79.1 | 88.4 | 6334 | No | 64.2 | 51.8 | 90.6 | 78.6 |
| | | | | | | | | Yes | 69.9 | 54.8 | 89.4 | 80.0 |
| Mean (Std) | - | 72.9 (1.1) | 32.0 (0.5) | | 78.8 (1.7) | 88.1 (1.1) | - | No | 66.4 (3.4) | 53.7 (3.6) | 92.2 (1.2) | 79.6 (1.5) |
| | | | | | | | | Yes | 76.1 (4.5) | 63.0 (5.4) | 92.7 (2.2) | 83.8 (2.3) |

**Table S7. Hyperparameters used for MaskRCNN and NuCLS model training.**

| MaskRCNN hyperparameter (refinement of noisy suggestions) | | NuCLS hyperparameters (learning from nucleus annotations) | |
|-----------------------------------------------------------|---------------|-----------------------------------------------------------|---------------------------|
| Backbone | Resnet50 | Backbone | Resnet18 |
| Pretraining | Imagenet | Pretraining | Imagenet |
| Input (cropped) image size | 128 x 128 | Input (cropped) image size | 300 x 300 |
| Max. ground truth nuclei per image | 30 | Scale factor | 4 |
| Max. detections per image (inference) | 200 | Scale factor jitter | 0.4 |
| Batch size | 8 | Max. ground truth nuclei per image | None |
| Optimizer | SGD | Max. detections per image (inference) | 300 |
| Learning rate | 1.00E-04 | Batch size | 4 |
| Momentum | 9.00E-01 | Optimizer | SGD |
| Length of anchor sides in pixels | 8,16,32,64,128 | Learning rate | 2.00E-03 |
| ROIs after NMS (training) | 500 | Momentum | 9.00E-01 |
| ROIs after NMS (inference) | 1000 | Length of anchor sides in pixels | Scale factor x (12,24,48) |
| NMS threshold for RPN proposals | 0.7 | ROIs after NMS (training) | 3000 |
| | | ROIs after NMS (inference) | 3000 |
| | | NMS threshold for RPN proposals | 0.7 |

**Table S8. List of interpretable features used as input for DTALE.** The features were extracted using the HistomicsTK package's *compute_nucleus_features* method; see digitalslidearchive.github.io.

| Category | N | Description | Feature | Category | N | Description | Feature |
|---|---|---|---|---|---|---|---|
| **Size** | 4 | Pixels occupied by the nucleus | Area | **Edges** | 8 | Gradients and canny edge filters associated with the hematoxylin intensity channel *(Zwillinger and Kokoska. CRC standard probability and statistics tables and formulae, Crc Press, 1999.)* | Mag.Mean |
| | | Length of major/minor axes of the ellipse with the same 2nd central moments | MajorAxis | | | | Mag.Std |
| | | | MinorAxis | | | | Mag.Skew |
| | | Pixelated perimeter using 4-connectivity | Perimeter | | | | Mag.Kurt. |
| **Shape** | 6 | Similarity to the shape of a circle | Circularity | | | | His.Entropy |
| | | Eccentricity of fitted ellipse (a measure of aspect ratio) | Eccentricity | | | | His.Energy |
| | | Diameter of a circle with the same area | Equiv.Diam. | | | | Canny.Sum |
| | | Ratio of nucleus area to its bounding box | Extent | | | | Canny.Mean |
| | | Aspect ratio of a fitted ellipse | Min.Maj.Axis | **Texture** *(Haralick et al., 1973.)* | 2 | Angular 2nd moment (ASM): A measure of homogeneity | Mean |
| | | A measure of convexity | Solidity | | | | Range |
| | 6 | Fourier shape descriptors. These represent simplifications of object shape. *(Zhang et al., ICIMADE01 2001)* | FSD1 | | 2 | Contrast: Intensity variation for neighbouring pixels | Mean |
| | | | FSD2 | | | | Range |
| | | | FSD3 | | 2 | Correlation: Intensity correlation for neighboring pixels | Mean |
| | | | FSD4 | | | | Range |
| | | | FSD5 | | 2 | Sum of squares: A measure of variance | Mean |
| | | | FSD6 | | | | Range |
| **Intensity** | 12 | Nucleus intensity features, extracted from the hematoxylin channel. *(Zwillinger and Kokoska. CRC standard probability and statistics tables and formulae, Crc Press, 1999.)* | Min | | 2 | Inverse difference moment: A measure of homogeneity | Mean |
| | | | Max | | | | Range |
| | | | Mean | | 4 | Sum average & Sum variance for all features | Mean |
| | | | Median | | | | Range |
| | | | MeanMed.Diff | | 2 | Sum entropy features | Mean |
| | | | Std | | | | Range |
| | | | IQR | | 2 | Entropy | Mean |
| | | | MAD | | | | Range |
| | | | Skewness | | 4 | Difference variance & Difference entropy | Mean |
| | | | Kurtosis | | | | Range |
| | | | HistEnergy | | 4 | Information Measure of Correlation (IMC) (2 types) | Mean |
| | | | HistEntropy | | | | Range |

# Supplementary figures

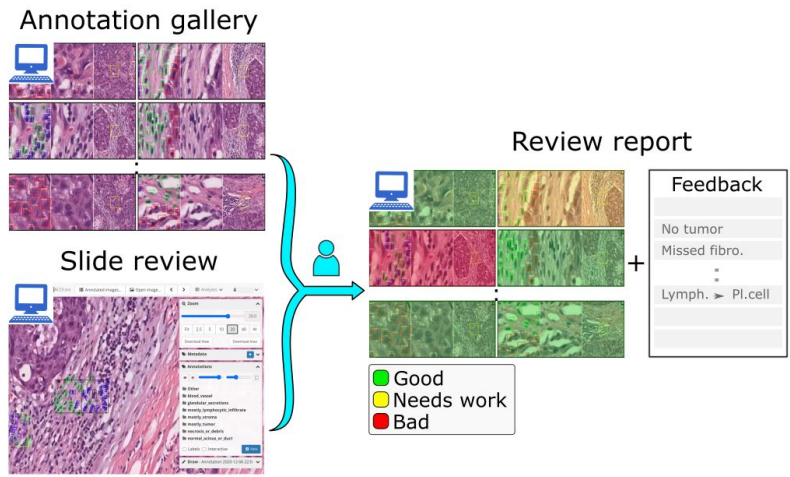

**Figure S1. Use of review galleries for scalable review of single-rater annotations.** Single-rater annotations were corrected by two study coordinators, in consultation with a senior pathologist. The pathologist was provided with a mosaic review gallery showing a bird's eye view of each FOV, with and without annotations, and at high and low power. The pathologist was asked to assign a per-FOV quality assessment. If the pathologist wanted further context, they were able to click on the FOV and pan around the full whole-slide image. They were also able to provide brief comments to be addressed by the coordinators, for eg. "change all to tumor". A demo is provided at the following video: https://youtu.be/Plh39obBg_0 .

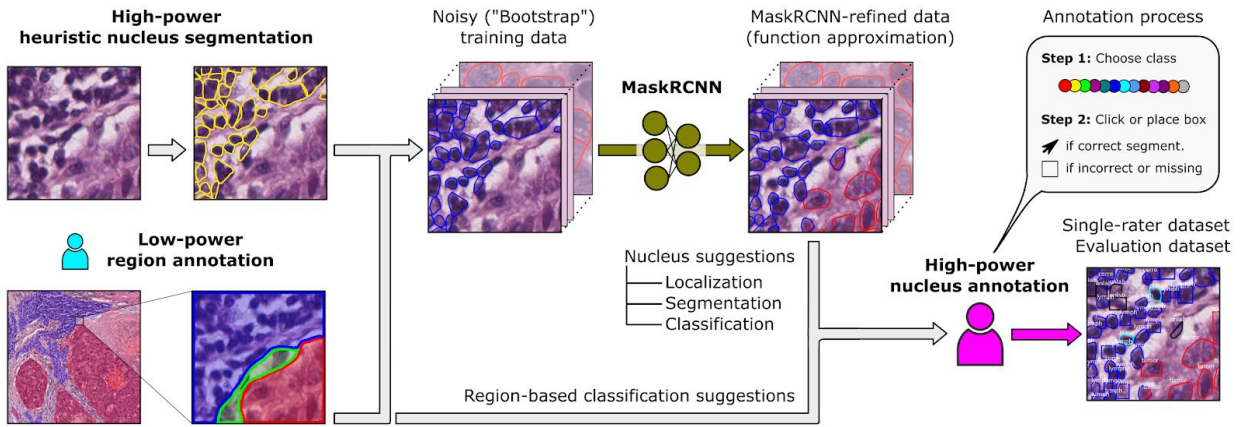

**Figure S2. Process for obtaining algorithmic suggestions for scalable assisted annotation.** Nucleus segmentation boundaries were derived using image processing heuristics at a high magnification. Low-power region annotations, approved by a practicing pathologist, were then used to assign an initial class to nuclei. This combination of noisy nuclear segmentation boundaries and region-derived classifications are the *bootstrapped* suggestions. These noisy algorithmic suggestions were the basis for annotating the Bootstrap control multi-rater dataset. A MaskRCNN model was then used as a function approximator to smooth out some of the noise in the bootstrapped suggestions. Participants were able to view these refined suggestions, along with low-power region annotations, when annotating the single-rater and Evaluation datasets.

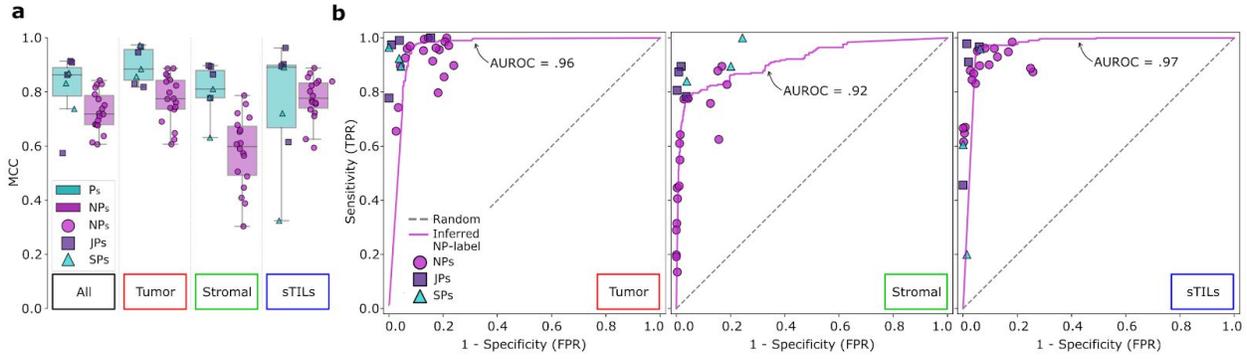

**Figure S3. Super-class accuracy of participant annotations and inferred NP-labels (Evaluation dataset).** The accuracy is measured against the inferred P-truth.

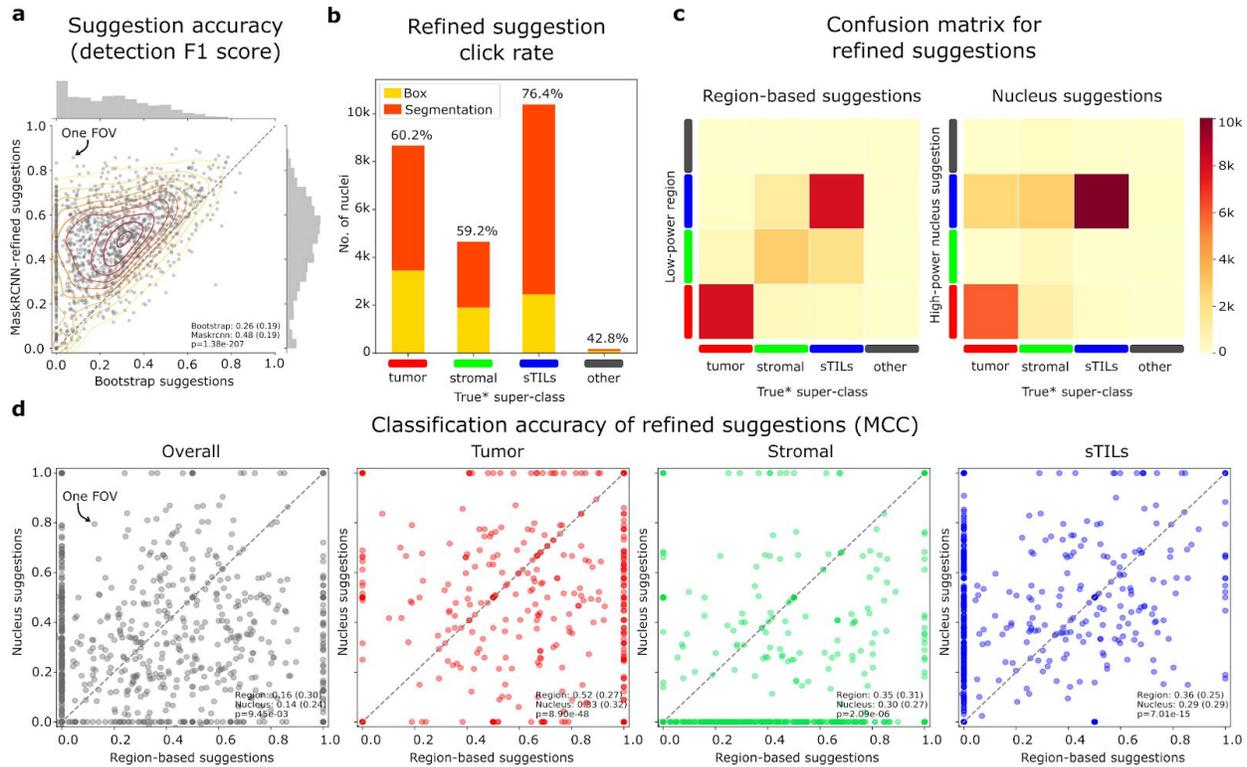

**Figure S4. Accuracy of algorithmic suggestions (single-rater dataset).** The accuracy is measured against the corrected single-rater dataset. **a.** Per-FOV detection accuracy of algorithmic data at the two stages of obtaining algorithmic suggestions; i.e. how well do the suggestions correspond to real nuclei? MaskRCNN refinement significantly improves suggestion accuracy. **b.** Number of MaskRCNN-refined suggestions that correspond to a segmentation (i.e. were clicked) or a bounding box. **c.** Concordance between suggested classes and classes assigned by participants. Region-based suggestions were, broadly-speaking, more concordant with the true classes, but nucleus suggestions had a higher recall for sTILs. **d.** Comparison of the classification accuracy (MCC) of low-power region class and high-power MaskRCNN-derived nucleus class. Note how region-based and nucleus-based suggestions have disparate accuracies for different FOVs and classes. Hence, there was value in providing the participants with both forms of suggestion.

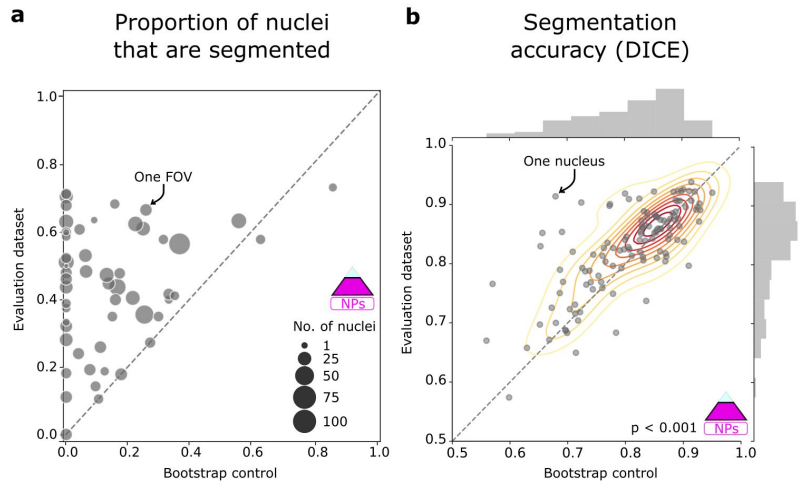

**Figure S5. Abundance and segmentation accuracy of clicked algorithmic suggestions (multi-rater datasets). a.** Proportion of nuclei in the FOV that were inferred to have good segmentation. Circle size represents the number of nuclei in that FOV. The proportion is notably higher for the Evaluation dataset than the Bootstrap control. **b.** Accuracy of algorithmic segmentation boundaries for nuclei that were inferred to have accurate segmentation boundaries in both the Evaluation dataset and Bootstrap control. The comparison is made against manual segmentations obtained for the same nuclei from one senior pathologist. Most clicked algorithmic segmentations were very accurate, and have a DICE coefficient above 0.8. The accuracy was slightly higher for MaskRCNN-refined suggestions.

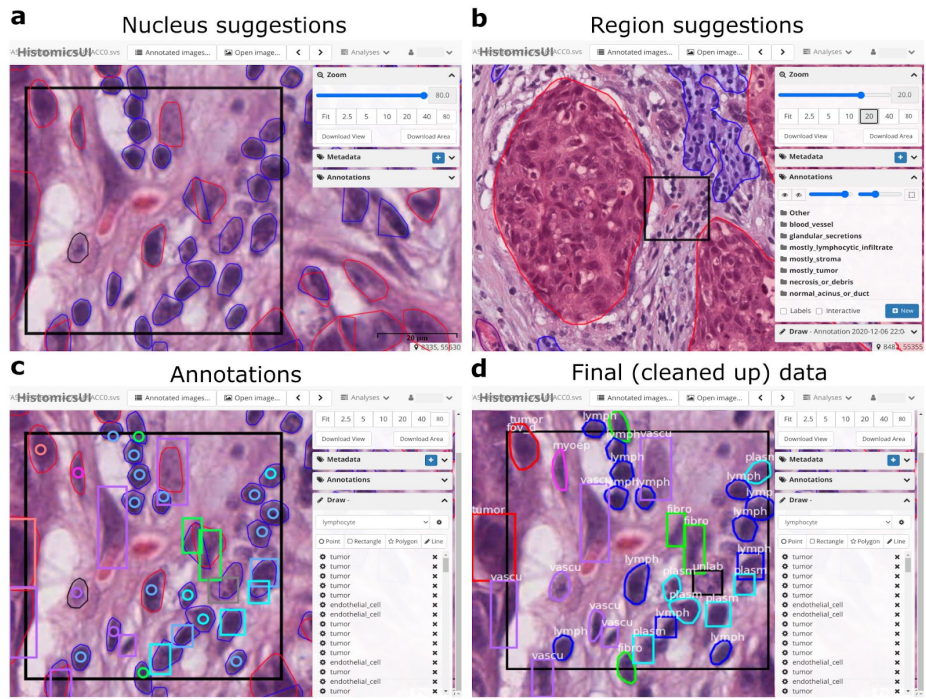

**Figure S6. Annotation procedure on HistomicsUI. a-b.** Participants were shown suggestions for nucleus segmentation boundaries, as well as two types of classification suggestions: low-power region suggestions and high-power nucleus classification suggestions. The FOV shown here is almost entirely present in a stromal region, but contains multiple scattered sTILs that were not dense enough to be captured as a sTILs "region". **c.** Participants' annotations were either points/clicks, for accurate segmentations, or bounding boxes. They picked the color/class of their annotations beforehand, and were told to simply ignore any inaccurate suggestions. Participants were able to turn the suggestions off for a clear view of the underlying tissue. **d.** Participant annotations and algorithmic suggestions were ingested into a database and processed to provide cleaned up data, which was then pushed for viewing on HistomicsUI for correction and review.

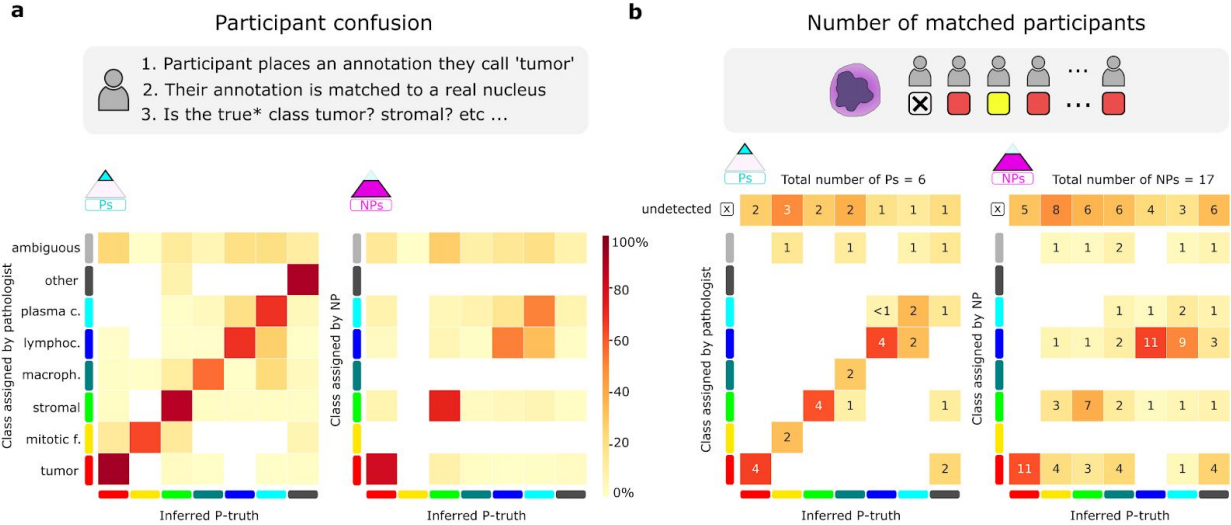

**Figure S7. Confusion matrix of participant annotations (Evaluation dataset). a.** Confusion of annotations placed by the participants, putting aside detection. Here, we ask the question, if a participant places an annotation they call tumor, and it matches a true nucleus, what is the class of that nucleus? By definition, there are no "ambiguous" true nuclei. **b.** For each true nucleus, how many of the participants detected it, and if so, what class did they assign? Note that since truth inference takes participant reliability into account, the inferred P-truth does not have to correspond to the most commonly assigned class. Empty entries are values <1.

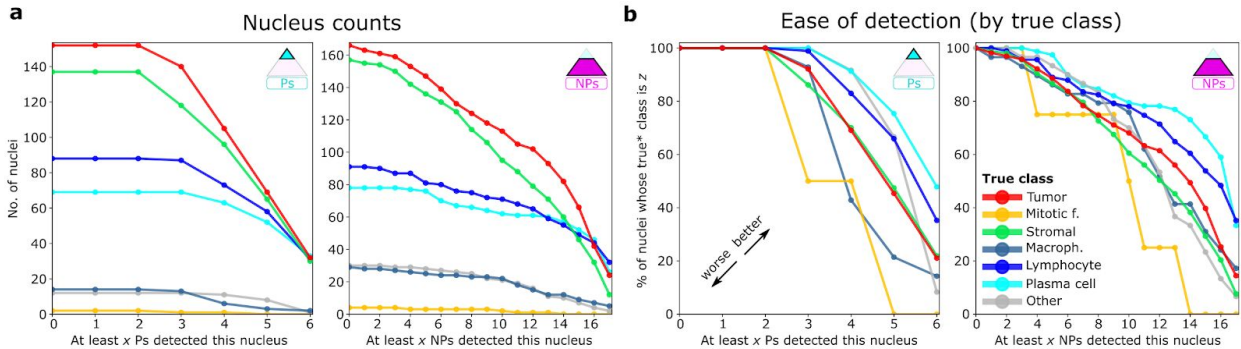

**Figure S8. Ease of detection of various nucleus classes (Evaluation dataset).** If we know for a fact this is, say, a lymphocyte, how many participants detected it, even if they called it something else?. True class is the inferred P-truth. The color coding used is explained in panel b. **a.** Nuclei counts, broken down by class and the number of matched participants. **b.** Ease of detection of nuclei by true class. Interpreting, say, the blue curve goes like this: 100% of lymphocytes were detected by at least 3 pathologists, ~80% were detected by 4 pathologists, and so on.

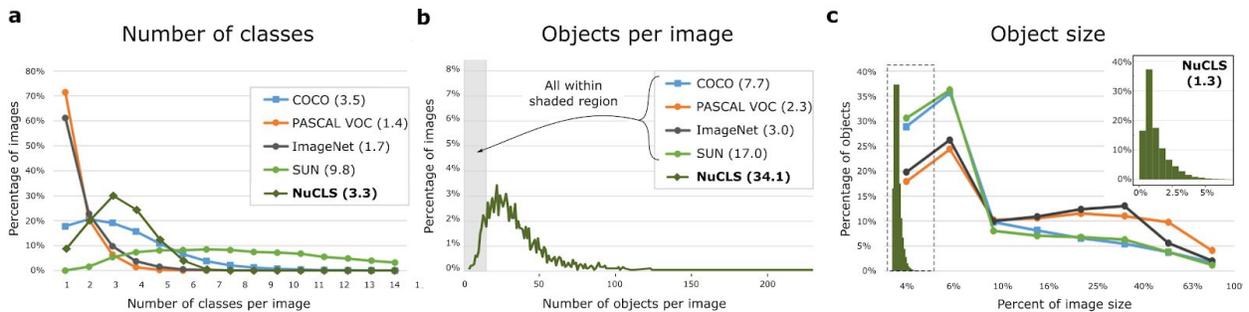

**Figure S9. Comparison of the corrected single-rater dataset with canonical "natural" object detection datasets.** The panels in this figure have been reproduced and modified with permission from: *Lin, Tsung-Yi, et al. "Microsoft COCO: Common objects in context." arXiv:1405.0312 [cs.CV].*

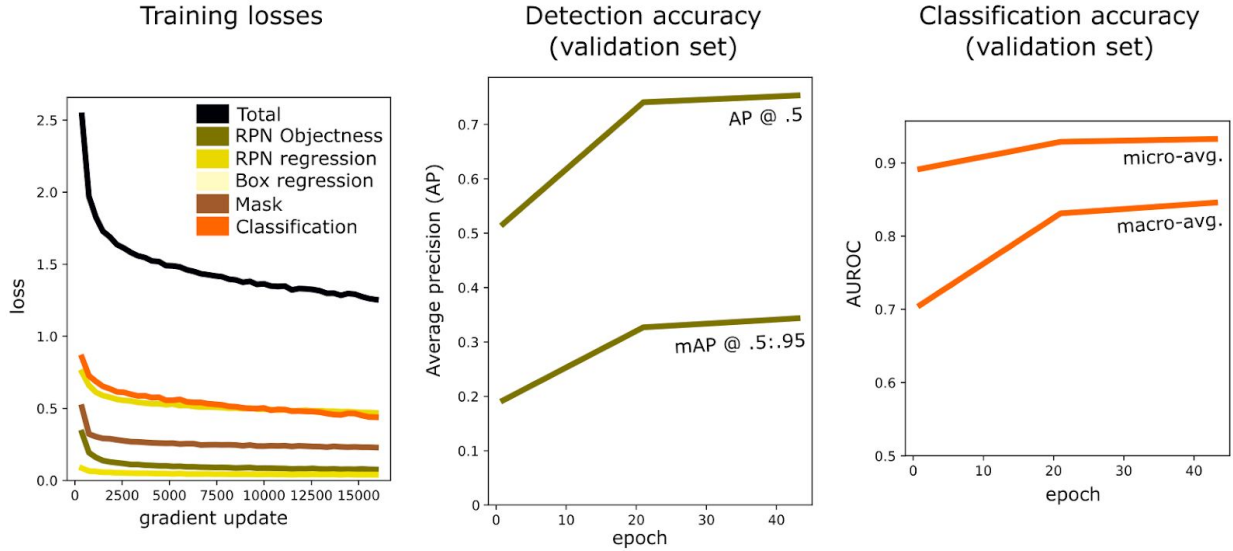

**Figure S10. Progression of NuCLS model training and convergence on fold 1.** Our prototyping experiments on fold 1 (not shown) showed that the detection model started overfitting after 15k detection updates, so we froze detection weights after 15k iterations and allowed 1k extra iterations for fine-tuning of the classification layers. Abbreviations: RPN, region proposal network; AP @ .5, average precision when a threshold of 0.5 is used for considering a detection to be true, mAP @ .5:.95, mean average precision at a range of detection thresholds between 0.5 and 0.95; AUROC, area under receiver-operator characteristics curve.

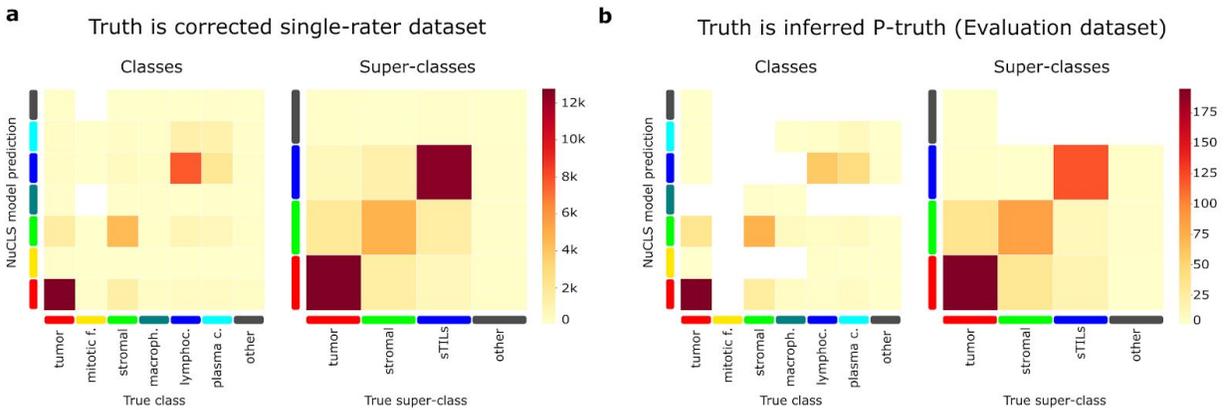

**Figure S11. Confusion matrix of NuCLS model predictions on the testing sets.** For each of folds 2-5, the NuCLS model trained on the corrected single-rater dataset training slides was used to predict FOVs from the corresponding testing set slides. The counts shown are aggregated over all testing sets. **a.** The corrected single-rater dataset is considered to be the truth. **b.** Inferred P-truth on the Evaluation dataset is considered to be the truth.

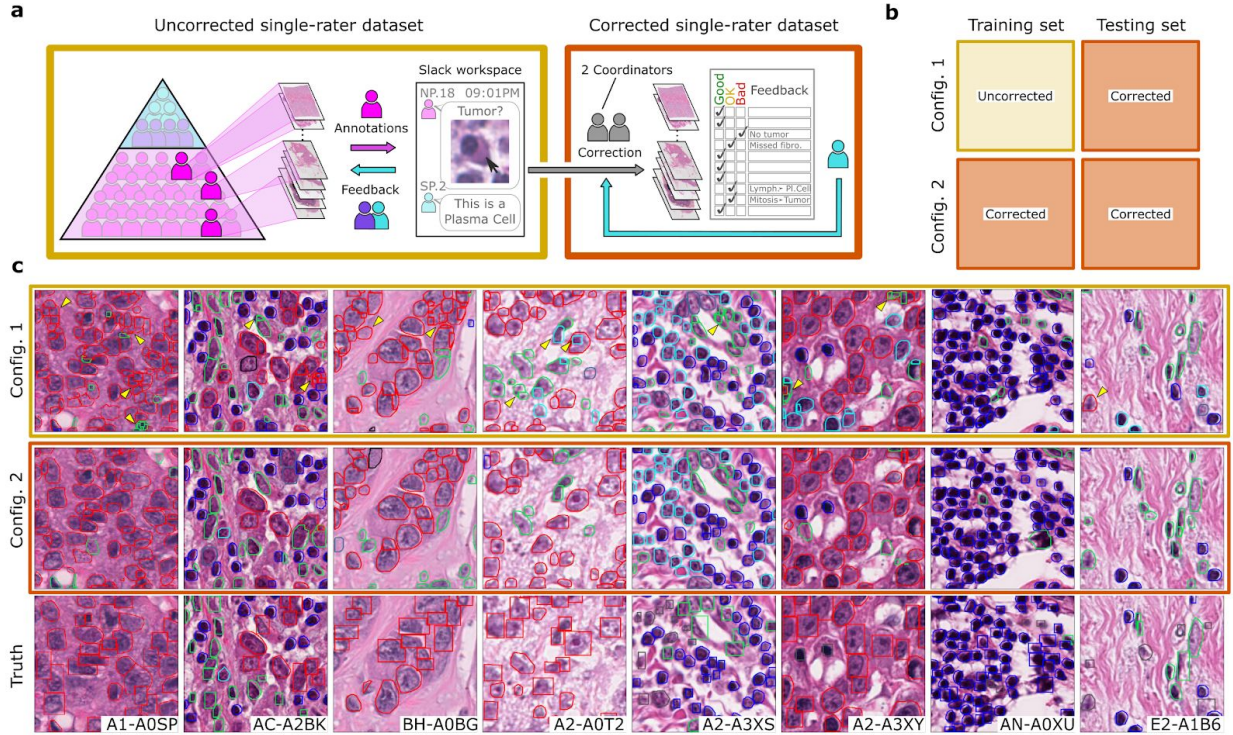

**Figure S12. Qualitative examination of the generalization accuracy of a NuCLS model trained on the uncorrected single-rater dataset.**

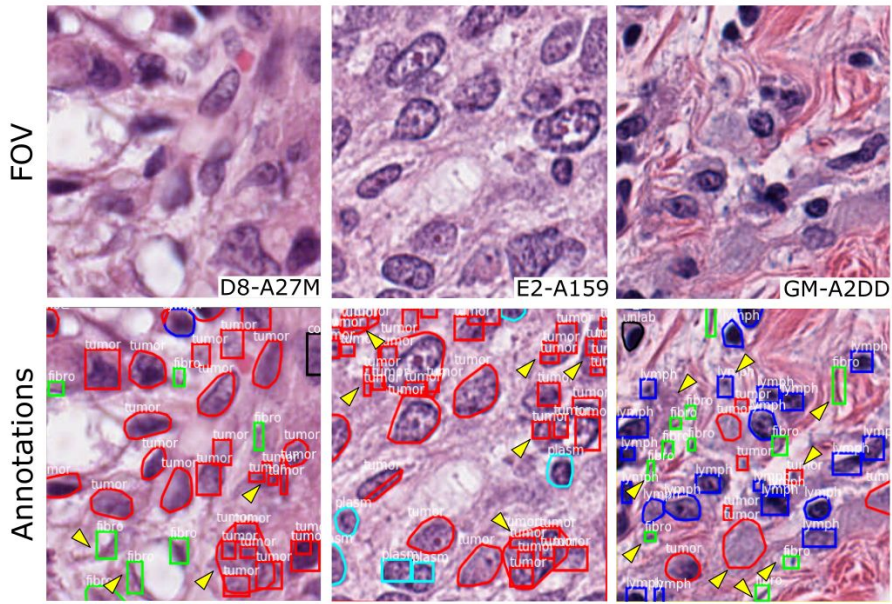

**Figure S13. Sample poor annotation data excluded during the single-rater dataset correction process.** Despite having received initial training and feedback, the NP who generated these annotations was confused about what is a nucleus, and frequently considered chromatin clumps or artifacts as nuclei (arrows).

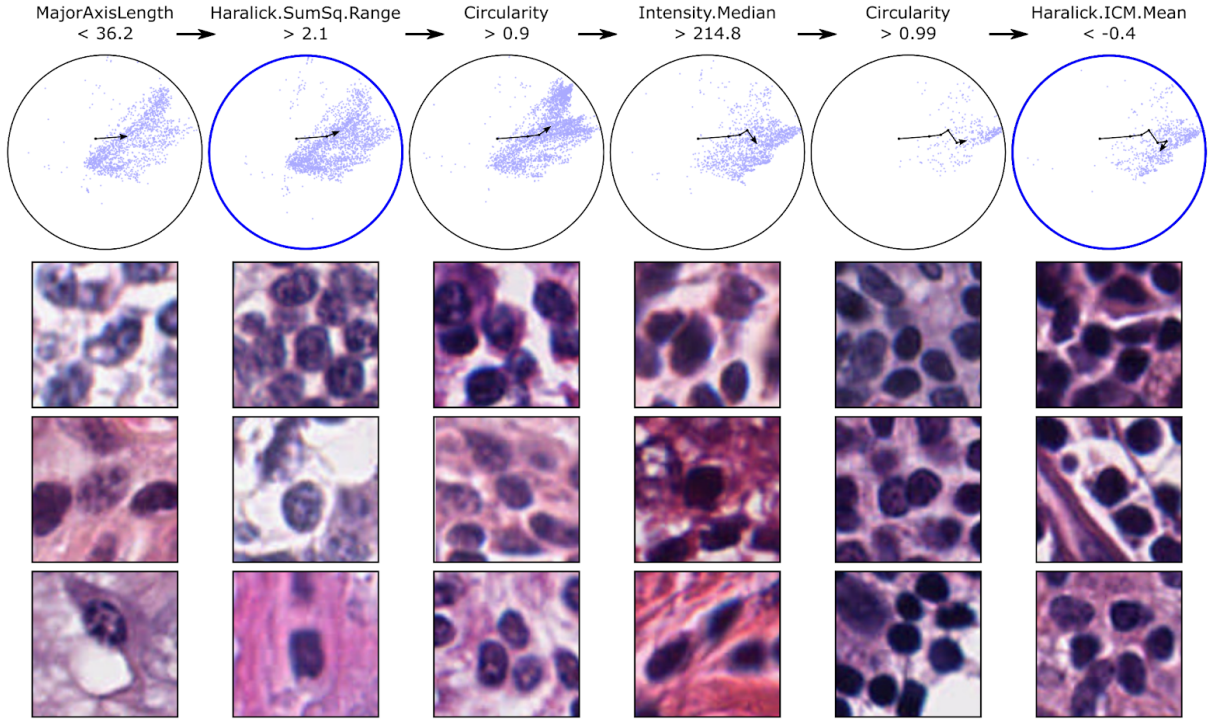

**Figure S14. DTALE enables fine-grained approximation of NuCLS model decisions.** Here we approximate the process by which NuCLS classifies nuclei as lymphocytes. The UMAP embedding is shown, along with an overlay of the DTALE path for lymphocyte classification. An intermediate node in the DTALE path corresponds to the most *representative* global explanation of NuCLS lymphocyte decisions (left blue circle). The initial set of decision criteria (MajorAxisLength < 36.2 and Haralick.SumSq.Range > 2.1) are our best global explanation for arriving at a lymphocyte classification (F1=0.74). When four extra decision criteria are met, we arrive at the most *discriminative* explanations (second blue circle). These criteria are highly specific to lymphocyte classifications (precision=0.98). In addition to providing global per-class explanations, DTALE also provides fine-grained, *within-class*, approximations of NuCLS decision-making. Because DTALE relies on regression trees, we can provide six explanations for different lymphocytes in our dataset, ranging along a spectrum from ambiguous to highly typical morphology.

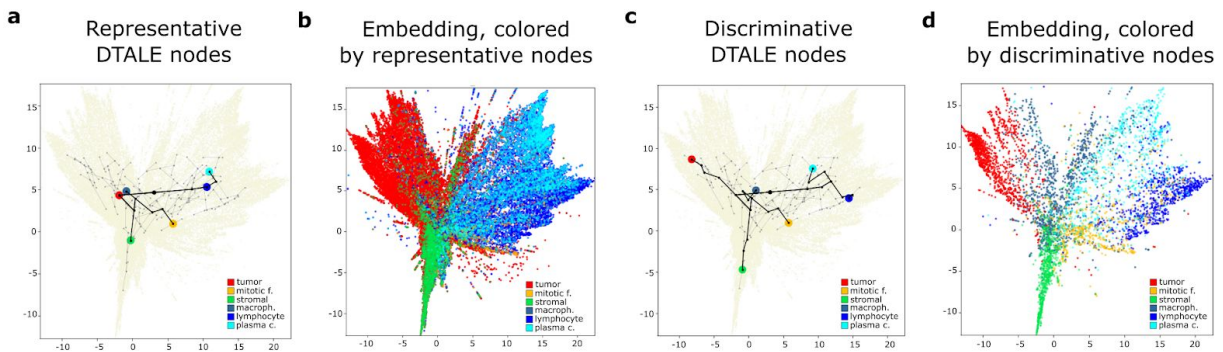

**Figure S15. Representative vs discriminative approximation of NuCLS model decisions using DTALE. a.** Overlay of the full DTALE tree (light gray) on top of the embedding to which it was fitted. In black, we show paths to the nodes that allow representative approximation of NuCLS decisions, i.e. highest F-1 score. **b.** Nuclei that correspond to representative DTALE nodes. **c.** DTALE nodes that correspond to the most discriminative approximation of the NuCLS decisions, i.e. highest precision. **d.** Nuclei that correspond to discriminative DTALE nodes.

**1** Do unconstrained agglomerative clustering with maximum linkage using bounding boxes from participants *{$P_1$, $P_2$, ...}*

**2** Cut at linkage threshold *1 - $t^*$* (where $t^*$ is the threshold IOU)

**3** For each cluster $C_i$ (corresponding to top-level node $N_i$)
  **3.1** For each "don't-link" set $S_j$
    **3.1.1** Check if more than one member in $S_j$ is present in $C_i$
    **3.1.2** For each extra member $S_{jk}$ in $C_i$
      **3.1.2.1** Check the next low-level node $N_{i-1}$
       > If there are no members from $S_j$ in $N_{i-1}$ :
         - If $C_{i-1}$ does not exist :
          → Set $N_{i-1}$ as a new cluster $C_{i-1}$
         - Assign $S_{jk}$ to $C_{i-1}$
       > Else :
         - Check the next low-level node $N_{i-2}$ (repeat 3.1.2.1)
      **3.1.2.2** If no nodes without members from $S_j$ found :
       > Assign $S_{jk}$ as a separate one-leaf cluster

**4** For each cluster $C_i$
  **4.1** Find IOU of bounding boxes of members {$C_{i1}$, $C_{i2}$, ..., $C_{iN}$}
  **4.2** Assign member $C_{im}$ = argmax(mean IOU) as the medoid

**Figure S16. Algorithm for obtaining anchor proposals through constrained agglomerative clustering.** We cluster bounding boxes from participants to get the *anchor proposals* corresponding to potential nucleus locations. Note that the threshold we use for maximum linkage, t*, is influential in determining how many anchors we get. We make sure that annotations from the same participant do not end up in the same cluster by creating sets of "don't-link" bounding boxes. The final anchor proposals are the anchor medoids; using medoids ensures that the box anchor proposals correspond to real nucleus boundaries.

## Supplementary file: Annotation protocol

Welcome to the breast cancer nucleus annotation project! The purpose of this project is to investigate a scalable data collection and refinement procedure, and to create a large-scale dataset for training and validation of machine learning algorithms.

**> Please view the introductory video before diving into this document.**
**> Please read this document in its entirety before making annotations.**

There are three categories of participants:

- **NP (Non-pathologist)** - Did not receive anatomical pathology residency training.
- **JP (Junior pathologist)** - Pathology residents with < 2 years of training.
- **SP (Senior pathologist)** - Attendings or pathology residents with > 2 years of training.

There are two required annotation assignments for each NP:

- **Single-rater dataset:** You can ask questions and receive feedback from pathologists.
- **Multi-rater datasets:** No feedback will be provided. Annotate to the best of your ability.

**> General remarks:**

- Use a **comfortable mouse, table and monitor**. This greatly impacts comfort and quality.

- When in doubt, take a screenshot and post a question on **Slack** for review & feedback.

- Remember, the algorithm is **learning** what we teach it (Garbage In → Garbage Out).

**> Annotation workflow:**

- <u>**Step 1:**</u> View the **region-level annotations.** These are the low-power classification suggestions.

- <u>**Step 2:**</u> Go to **medium power** (20x) and **reduce transparency**. Examine the underlying tissue.

- <u>**Step 3:**</u> Zoom on the FOV at **maximum power** (40x or 80x, depending on slide).

- <u>**Step 4:**</u> **Start annotating**. The process is illustrated in the introductory video. Briefly, the steps are:

    > Pick an annotation class/color (feel free to rely on or ignore algorithmic suggestions)

    > If an algorithmic boundary is correct, place a dot.

    > Otherwise, place a bounding box around the nucleus.

**> Specific annotation rules:**

- Only annotate the Fields-of-View (FOVs) that were picked for you.

- If a nucleus extends beyond the FOV boundary, make sure your bounding box covers its full extent (i.e. extend your rectangle outside the FOV as well).

- Make sure each FOV is complete before moving to the next. Missing annotations may confuse our algorithms and make validation difficult!

- Make sure to annotate **in this order**: **Single-rater dataset → Multi-rater dataset 1 → Multi-rater dataset 2 and/or Multi-rater dataset 3.** SPs and JPs do not have a single-rater dataset (but they *do* have multi-rater datasets). Pathologists are kindly asked to respond to questions on Slack.

> ***Explanatory note:*** *We asked the participants to annotate the single-rater dataset first because this also acted as their de-facto training, and they received feedback and could ask questions. The multi-rater datasets were blinded to avoid biasing the participants. Multi-rater dataset 1 is the unbiased control dataset (no algorithmic suggestions), and was annotated first for the same reason.*

- After you annotate your **first FOV**, take a screenshot and share it on the **Slack group** to get **approval & feedback before continuing.** This acts as a test of your understanding.

- After every slide in the single-rater dataset, please ask for feedback from the SPs and/or study coordinator. Do not post a screenshot of every single FOV, simply post the slide ID on the group and SPs/coordinator will go to the slide and make suggestions/corrections where necessary.

- Share a screenshot of anything that you are unsure of, making sure to also share the slide name so that the SPs and study coordinators can take a closer look at various magnifications. Nuclei are often vague. If you are unsure about the class of a nucleus, either:
  - Ask what it is on the group and receive feedback from SP (preferred).
  - Assign is the class *unlabeled*.

  Make as much effort to classify nuclei as possible; only use the *unlabeled* class in a minority of cases.

- Make sure the **bounding box** is **tight** around the nucleus **NOT** the entire cell.

- Do not trust the computer suggestions too much. If the algorithmic boundaries are just slightly off then it's OK, otherwise use a bounding box instead.

- Never rotate the slide before annotating. All boxes should have the same orientation as the FOV.

**> Notes about specific annotation classes:**

*Notes that address some frequently asked questions on the Slack group.*

- **Tumor:** Malignant cells are very heterogeneous in shape. They tend to have hyperchromatic, eccentric nuclei, and tend to be crowded and irregular. See any standard pathology textbook.

- **Fibroblasts:** Stromal nuclei tend to be elongated and shaped like a cigar. May also have a rounder shape. The tell-tale sign is their presence in stroma in alignment with the collagen fibres. Some fibroblasts close to the tumor may be **activated** (i.e. have tumor-like morphology).

- **Lymphocytes:** Small, round, condensed, central nucleus. Tend to be grouped together.

- **Plasma cells:** May confuse with lymphocytes. Plasma cells are less common than lymphocytes; when in doubt, ask on Slack. They tend to have an eccentric, large, textured nucleus (described as *cart-wheel*, but rarely seen as such) with a pale perinuclear halo. Also tend to have eosinophilic cytoplasm

- **Macrophages:** Usually difficult to ascertain. They tend to be larger than lymphocytes, sometimes have vacuolated or frothy cytoplasm, have thin round-to-uniform (bean shaped) nuclei with variable nucleoli.

**> Troubleshooting:**

| # | Situation | How to handle | Examples |
|---|---|---|---|
| **1** | Annotations take a long time to load. | - Close all programs running in the background.<br>- Close all other Google Chrome tabs, especially videos (except this document, which you always have to refer to)<br>- Switch from tablet to computer<br>- Switch to a faster internet connection | N/A |
| **2** | Algorithm correctly predicts both nucleus boundary and class | Place a dot inside the nucleus with the correct class | 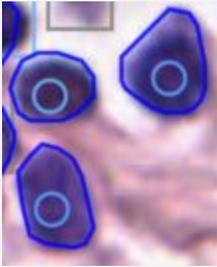 |
| **3a** | Algorithm correctly predicts nucleus boundary but assigns incorrect class, and you know the correct class | - If you know correct class: Place a dot with correct class inside the nucleus | 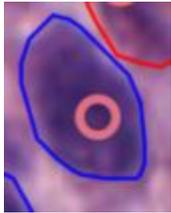 |

| 3b | Same as 3a, but you do not know the correct class | - If you are new to this or are in doubt, take a snapshot and ask for pathologist feedback.<br>- If you are confident in your ability (eg you have been annotating many FOVs or are a pathologist), i.e. the nucleus is vague and cannot be classified using just H&E: place a dot with the class *unlabeled*. | |
| --- | --- | --- | --- |
| 4 | Algorithm incorrectly predicts nucleus boundary or completely misses the nucleus | Place a rectangle with the correct class and color around the nucleus. The rectangle must be tight (i.e. it should be precise, not too large or too small). | 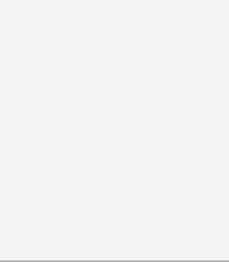 |
| 5 | The algorithm clumps multiple nuclei together | Place a rectangle around each nucleus and ignore the algorithmic suggestion. | 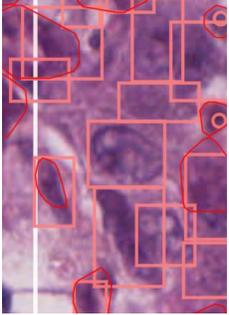 |
| 6a | A nucleus extends beyond the edge of the FOV and I need to place a <u>dot</u> | Place the dot inside the FOV. | 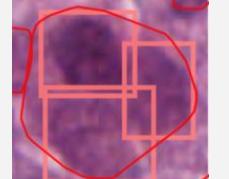 |
| 6b | A nucleus extends beyond the edge of the FOV and I need to place a <u>rectangle</u> | Extend your rectangle to encompass the full extend of the nucleus | 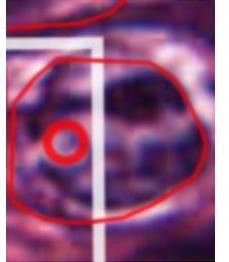 |
| 7 | I cannot see the underlying tissue | Reduce the annotation transparency | |
| 8 | I know the nucleus class but it is not in the | - If you are an NP: ask on Slack; a pathologist may recommend a class.<br>- If you are a pathologist, create your own | |

| | standard classes | class, and notify the study coordinator. | |
|---|---|---|---|
| **9** | The algorithm predict two overlapping boundaries for the same nucleus; only one is correct | If it is possible to place the dot inside the inside the correct boundary, but outside the incorrect one, do so. | 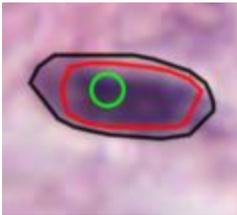 |
| **10** | There is necrotic debris or collagen | Ignore it. Do **not** annotate debris or non-nuclear material. | 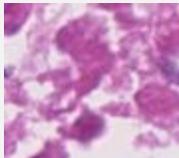 |
| **11** | There are red blood cells | Ignore. Do **not** annotate RBCs | 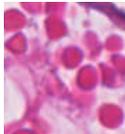 |
| **12** | There is a multinucleated giant cell or a cell-eat-cell phenomenon (cannibalism) | Classify each nucleus independently. We operate at the level of nuclei, not cells, in this project. | 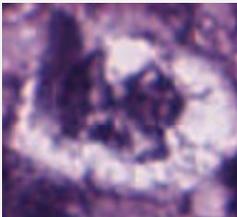 |
| **13** | There are overlapping nuclei and the bounding boxes will have to overlap to capture full extent. | No problem; use overlapping bounding boxes in this case. | 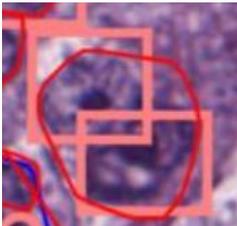 |
| **14** | The nuclei are very textured and have prominent nucleoli | Don't be fooled!! Malignant nuclei can have a very textured appearance and prominent nucleoli so you may think they are multiple nuclei but are one nucleus!! By the way, in the image below, there are many vacuoles that were mistaken as being nuclei. This is a vacuolated phenotype. | |

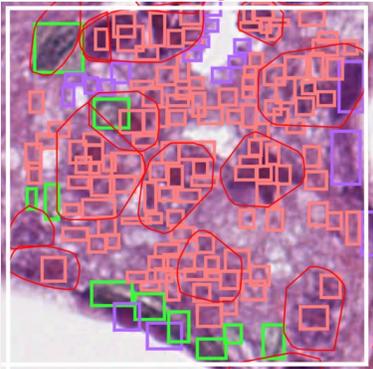

| | | | |
|---|---|---|---|
| **15** | The slide is quite difficult; stroma is difficult to distinguish from tumor. | Make sure you follow step 1 in the *Annotation workflow* section. Anything outside tumor regions may still be a tumor nucleus, but is more likely to be a fibroblast, lymphocyte, plasma cell etc. | 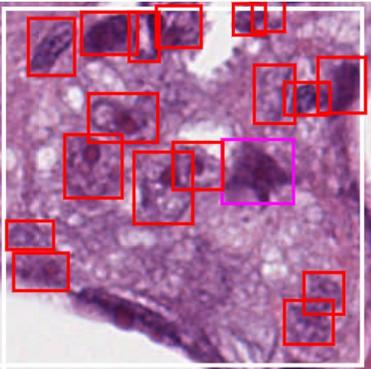 |
| **16** | After finishing many FOVs, I discovered (or was told) that I have a systematic error in classifying nuclei (eg. all plasma cells mistakenly called tumor) | Notify one of the study coordinators and we will run a program (python script) to do this automatically for you. | |